\newcommand{\bea}{\bgroup\setlength{\arraycolsep}{0pt}\begin{eqnarray}}
\newcommand{\eea}{\end{eqnarray}\egroup}
\newcommand{\half}{\mbox{$\frac{1}{2}$}}
\newcommand{\ihalf}{\mbox{$\frac{i}{2}$}}
\newcommand{\br}{\nonumber\\ &&\hbox{}}
\newcommand{\lp}{\loarrow{\bm{\partial}}}
\newcommand{\rp}{\roarrow{\bm{\partial}}}
\documentclass[draft,amsmath,aps,prd,preprint,eqsecnum,groupedaddress,%
nofootinbib]{revtex4}
\usepackage{bm}
\begin{document}
\preprint{ANL-HEP-PR-02-031}

\title{Order-$\bm{v^4}$ Corrections to $\bm{S}$-wave Quarkonium Decay} 

\author{Geoffrey T.~Bodwin}
\author{Andrea Petrelli}
\altaffiliation{Present Address: IntesaBci, Risk Management Department, 
Capital Markets Division, Via Verri 4, 20121 Milano, Italy.
}
\affiliation{High Energy Physics Division, Argonne National Laboratory,
Argonne, IL 60439}

\date{May, 2002}

\begin{abstract}

We compute corrections of relative order $v^4$ to the rates for the
decays of ${}^1S_0$ heavy quarkonium into two photons and into light
hadrons and for the decays of ${}^3S_1$ heavy quarkonium into a lepton
pair and into light hadrons. In particular, we compute the coefficients
of the decay operators that have the same quantum numbers as the heavy
quarkonium. We also confirm previous calculations of the order-$v^2$
corrections to these rates. We find that the $v$ expansion converges
well for the decays of ${}^1S_0$ heavy quarkonium and for the decay of
${}^3S_1$ heavy quarkonium into a lepton pair. Large higher-order-in-$v$
corrections appear in the decay of ${}^3S_1$ heavy quarkonium into light
hadrons. However, we find that the series of coefficients of operators
with ${}^3S_1$ quantum numbers, which yields a large correction in order
$v^2$, yields a smaller correction in order $v^4$.

\end{abstract}

\maketitle

\section{Introduction}

A formalism for the first-principles calculation of heavy-quarkonium
decay rates in quantum chromodynamics (QCD) has been given in
Ref.~\cite{bbl}. This formalism is based on the effective field theory
nonrelativistic quantum chromodynamics (NRQCD). In it, one can write the
decay rate of a quarkonium state $H$ as
\begin{equation}
\Gamma(H)=\sum_n{F_n\over m^{d_n-4}}
\langle H|{\cal O}|H\rangle,
\end{equation}
where $F_n$ is a perturbatively calculable short-distance
coefficient, $m$ is the heavy-quark mass, the ${\cal O}_n$ are
four-fermion NRQCD operators, and $d_n$ is the mass dimension of ${\cal
O}_n$. The terms in the sum over $n$ may be classified according to
their orders in $v$ (Ref.~\cite{bbl}), where $v$ is the
heavy-quark--antiquark relative velocity. For charmonium, $v^2\approx
0.3$; for bottomonium, $v^2\approx 0.1$.

We concern ourselves in this paper with the decay of ${}^1S_0$
quarkonium into light hadrons and the decays of ${}^3S_1$ quarkonium
into lepton pairs and into light hadrons. The coefficients of the
operators of leading order in $v$ and of relative order $v^2$ have been
computed previously
\cite{bbl,Barbieri:1979be,Hagiwara:1980nv,Mackenzie:1981sf,Keung:jb,%
harris-brown,Barbieri:1975ki,Celmaster:1978yz,%
Beneke:1997jm}. Some of the coefficients of the order-$v^2$ operators
are sufficiently large as to cast doubt on the convergence of the $v$
expansion for charmonium and bottomonium. In particular, the order-$v^2$
correction to the rate for the decay of ${}^3S_1$ quarkonium into light
hadrons is $-5.32 \langle v^2 \rangle$, where $\langle v^2 \rangle$ is
the ratio of the expectation values of the order-$v^2$ and order-$v^0$
operators in the quarkonium state. Hence, in the case of charmonium, the
order-$v^2$ correction is more than 100\%.

In this paper, we compute the short-distance coefficients of the decay
operators, through order $v^4$, that have the same quantum numbers as
the quarkonium. Our calculations confirm previous results for the
short-distance coefficients of the order-$v^2$ operators. We find that
the $v$ expansion is well behaved for the decays of ${}^1S_0$ quarkonium
and for the decay of ${}^3S_1$ quarkonium into lepton pairs. In the case
of the decay of ${}^3S_1$ quarkonium into light hadrons, large
coefficients are associated with some of the operators of higher order
in $v$. For the operators with ${}^3S_1$ quantum numbers, a large
correction to the decay rate appears in order $v^2$, but the correction
in order $v^4$ is considerably smaller. This suggests that the $v$
expansion for operators with a given quantum number may converge well
once one goes beyond the first nontrivial order.

\section{NRQCD Decay rates}

In this section, we present the NRQCD factorization expressions for the
rates of ${}^1S_0$ quarkonium ({\it e.g.} $\eta_c$ or $\eta_b$) decay to
light hadrons (LH), ${}^3S_1$ quarkonium ({\it e.g.} $J/\psi$ or
$\Upsilon$) decay to light hadrons, ${}^1S_0$ quarkonium decay to two
photons, and ${}^3S_1$ quarkonium decay to $e^+e^-$.

Through relative order $v^4$, the rate for the decay of a ${}^1S_0$ 
state into light hadrons is given by
\begin{eqnarray}
\Gamma({}^1S_0 \to {\rm LH})
&=& {F_1({}^1S_0) \over m^2} 
        \langle {}^1S_0| {\cal O}_1({}^1S_0) |{}^1S_0\rangle
+{G_1({}^1S_0) \over m^4}
        \langle {}^1S_0| {\cal P}_1({}^1S_0) |{}^1S_0\rangle\nonumber\\
&&\hbox{}+{F_8({}^3S_1) \over m^2}
        \langle {}^1S_0| {\cal O}_8({}^3S_1) |{}^1S_0\rangle
+{F_8({}^1S_0) \over m^2}
        \langle {}^1S_0| {\cal O}_8({}^1S_0) |{}^1S_0\rangle\nonumber\\
&&\hbox{}+{F_8({}^1P_1) \over m^4}
        \langle {}^1S_0| {\cal O}_8({}^1P_1) |{}^1S_0\rangle
+{H_1^1({}^1S_0) \over m^6}
        \langle {}^1S_0| {\cal Q}_1^1({}^1S_0) |{}^1S_0\rangle\nonumber\\
&&\hbox{}+{H_1^2({}^1S_0) \over m^6}
        \langle {}^1S_0| {\cal Q}_1^2({}^1S_0) |{}^1S_0\rangle.
\label{eta-decay}
\end{eqnarray}
The operators appearing in Eq.~(\ref{eta-decay}) are defined by
\begin{subequations}
\label{1s0-ops}
\begin{eqnarray}
{\cal O}_1({}^1S_0)&=&\psi^\dagger\chi\chi^\dagger\psi,\\
{\cal P}_1({}^1S_0)&=&{1\over 2}\left[\psi^\dagger\chi\chi^\dagger
(-\ihalf \tensor{\bf D})^2\psi+\psi^\dagger(-\ihalf \tensor{\bf D})^2
\chi\chi^\dagger\psi\right],\\
{\cal O}_8({}^3S_1)&=&\psi^\dagger\bm{\sigma}T_a\chi\cdot
\chi^\dagger\bm{\sigma}T_a\psi,
\label{o83s1}\\
{\cal O}_8({}^1S_0)&=&\psi^\dagger T_a\chi\chi^\dagger 
T_a\psi,\label{o81s0}\\
{\cal O}_8({}^1P_1)&=&\psi^\dagger(-\ihalf\tensor{\bf D})T_a
\chi\cdot\chi^\dagger(-\ihalf\tensor{\bf D})T_a\psi,\\
{\cal Q}_1^1({}^1S_0)&=&\psi^\dagger(-\ihalf\tensor{\bf D})^2\chi
\chi^\dagger(-\ihalf\tensor{\bf D})^2\psi,\\
{\cal Q}_1^2({}^1S_0)&=&{1\over 2}\left[\psi^\dagger\chi
\chi^\dagger(-\ihalf\tensor{\bf D})^4\psi
+\psi^\dagger
(-\ihalf\tensor{\bf D})^4\chi
\chi^\dagger\psi\right],\\
{\cal Q}_1^3({}^1S_0)&=&{1\over 2}\left[\psi^\dagger\chi\chi^\dagger
(\tensor{\bf D}\cdot g{\bf E}+g{\bf E}\cdot \tensor{\bf D})\psi
-\psi^\dagger(\tensor{\bf D}\cdot g{\bf E}+g{\bf E}\cdot\tensor{\bf D})\chi
\chi^\dagger\psi\right],
\end{eqnarray}
\end{subequations}
where the subscript $1$ or $8$ indicates that the operator is a color
singlet or a color octet, the superscript labels the three dimension-10
operators, $\psi$ is the Pauli-spinor field that annihilates a heavy
quark, $\chi^\dagger$ is the Pauli-spinor field that annihilates a heavy
antiquark, $D^\mu=\partial^\mu+igA^\mu$ is the gauge-covariant
derivative, $A$ is the ${\rm SU}(3)$-matrix-valued gauge field, $g$ is
the QCD coupling constant, $E^i=G^{0i}$, where $G^{\mu\nu}=\partial ^\mu
A^\nu-\partial^\nu A^\mu +ig[A^\mu,A^\nu]$ is the gluon field strength,
and the $\sigma^i$ are Pauli matrices. The operator $\tensor{\bf D}$ is
defined by $\chi^\dagger\tensor{\bf D}\psi=\chi^\dagger({\bf
D}\psi)-({\bf D}\chi)^\dagger\psi$. The relative signs of the
terms in each of these operators (and, in particular, ${\cal Q}_1^3$) are
fixed by the requirements of hermiticity and charge-conjugation
invariance.

The matrix element of ${\cal Q}_1^3$ does not appear in
Eq.~(\ref{eta-decay}) because, as we show in
Appendix~\ref{app:relation}, it can be eliminated in favor of ${\cal
Q}_1^1$ and ${\cal Q}_1^2$ through the use of the equations of motion.
From the velocity scaling rules in Ref.~\cite{bbl}, we find that, in the
${}^1S_0$ state, the operator ${\cal O}_1({}^1S_0)$ has a matrix element
of relative order  $v^0$, the operator ${\cal P}_1({}^1S_0)$ has a
matrix element of relative order  $v^2$, the operator ${\cal
O}_8({}^3S_1)$ has a matrix element of relative order  $v^3$, and the
operators ${\cal O}_8({}^1S_0)$, ${\cal O}_8({}^1P_1)$, ${\cal
Q}_1^1({}^1S_0)$, ${\cal Q}_1^2({}^1S_0)$, and ${\cal Q}_1^3({}^1S_0)$
have matrix elements of relative order  $v^4$. 

The contributions of order $\alpha_s^2$ and order $\alpha_s^3$ to the
short-distance coefficient $F_1({}^1S_0)$ have been computed in
Refs.~\cite{Barbieri:1979be,Hagiwara:1980nv} and are given in
Ref.~\cite{bbl}:
\begin{equation}
F_1({}^1S_0)= 
{\pi C_F \over N_c} \alpha_s^2(2m)
\Bigg\{ 1 + \left[ \left( {\pi^2 \over 4} - 5 \right) C_F
        + \left( {199 \over 18} -  {13 \pi^2 \over 24} \right) C_A
        - {8 \over 9} n_f \right] {\alpha_s \over \pi} \Bigg\},
\label{fetaNLO}
\end{equation}
where $N_c=3$ is the number of colors, $C_F=(N_c^2-1)/(2N_c)=4/3$, and
$C_A=N_c$. The contribution of order $\alpha_s^2$ to $G_1({}^1S_0)$ has
been computed in Refs.~\cite{Keung:jb,bbl}.\footnote{Short-distance
coefficients can be extracted from the results in Ref.~\cite{Keung:jb} by
first making the substitution $1/M^2_{\rm meson}\rightarrow
(1/4m^2)(1-\varepsilon/m)$, where $-\varepsilon$ is the binding energy,
and then making the identification
$\varepsilon/m\rightarrow \langle {}^1S_0| {\cal P}_1({}^1S_0)
|{}^1S_0\rangle/[m^2\langle {}^1S_0| {\cal O}_1({}^1S_0) 
|{}^1S_0\rangle]
\approx \langle {}^3S_1| {\cal P}_1({}^3S_1)
|{}^3S_1\rangle/[m^2\langle {}^3S_1| {\cal O}_1({}^3S_1)
|{}^3S_1\rangle]$.}
It is 
\begin{equation}
G_1({}^1S_0)=-{4\pi C_F\over 3N_c}\alpha_s^2.
\end{equation}
We note that, to leading order in $\alpha_s$, $[G_1({}^1S_0)]
/[F_1({}^1S_0)]=-4/3$. Hence, the first relativistic correction is
sizable in the case of the $\eta_c$. The contributions of order
$\alpha_s^2$ and order $\alpha_s^3$ to $F_8({}^3S_1)$ and $F_8({}^1S_0)$
have been computed by Petrelli {\it et al.} \cite{Petrelli:1998ge}:
\begin{subequations}
\begin{eqnarray}
F_8({}^3S_1)&=&{\pi n_f\over 3}\alpha_s^2(\mu)
\bigg\{1+{\alpha_s\over\pi}\bigg[-{13\over 4}C_F+
\bigg({133\over 18}+{2\over 3}\log 2-{\pi^2\over 4}\bigg)C_A-{10\over 
9}n_fT_F\nonumber\\
&&\qquad\qquad\qquad\hbox{}+2b_0\log{\mu\over 2m}\bigg]\bigg\}
+5\alpha_s^3\bigg(-{73\over 4}+{67\over 36}\pi^2\bigg),\label{f83s1}\\
F_8({}^1S_0)&=&2\pi B_F\alpha_s^2(\mu)
\bigg\{1+{\alpha_s\over\pi}\bigg[\bigg(-5+{\pi^2\over 4}\bigg)C_F+
\bigg({122\over 9}-{17\over 24}\pi^2\bigg)C_A-{16\over
9}n_fT_F\nonumber\\
&&\qquad\qquad\qquad\hbox{}+2b_0\log{\mu\over 2m}\bigg]\bigg\},
\label{f81s0}
\end{eqnarray}
\end{subequations}
where $\mu$ is the QCD renormalization scale, $n_f$ is the number
of light-quark flavors, $B_F=(N_c^2-4)/(4N_c)=5/12$, $T_F=1/2$, and
$b_0=(11/6)C_A-(2/3)T_Fn_f$. The contribution of order $\alpha_s^2$ to 
$F_8({}^1P_1)$ can be deduced from the results in Appendix~A~2 of 
Ref.~\cite{bbl}:
\begin{equation}
F_8({}^1P_1)={\pi N_c\over 6}\alpha_s^2.
\end{equation}

Owing to energy conservation, the operators associated with the
short-distance coefficients $H_1^1({}^1S_0)$ and $H_1^2({}^1S_0)$ cannot
be distinguished from each other in the Born-level decay of on-shell
quarks. Consequently, if one uses on-shell matching between NRQCD and
full QCD to compute the short-distance coefficients in Born-level decay
processes, one can compute only $H_1^1({}^1S_0)+H_1^2({}^1S_0)$, not the
individual coefficients. It is the quantity
$H_1^1({}^1S_0)+H_1^2({}^1S_0)$ that we compute in this paper.

Through relative order $v^4$, the decay rate for a ${}^3S_1$ state
into light hadrons is
\begin{eqnarray}
\Gamma({}^3S_1 \to {\rm LH})
&=& {F_1({}^3S_1) \over m^2} \,
        \langle {}^3S_1| {\cal O}_1({}^3S_1) |{}^3S_1\rangle
+{G_1({}^3S_1) \over m^4}
        \langle {}^3S_1| {\cal P}_1({}^3S_1) |{}^3S_1\rangle\nonumber\\
&&\hbox{}+{F_8({}^1S_0) \over m^2}
        \langle {}^3S_1| {\cal O}_8({}^1S_0) |{}^3S_1\rangle
+{F_8({}^3S_1) \over m^2}
        \langle {}^3S_1| {\cal O}_8({}^3S_1) |{}^3S_1\rangle\nonumber\\
&&\hbox{}+\sum_{J=0,1,2} {F_8({}^3P_J) \over m^4}
        \langle {}^3S_1| {\cal O}_8({}^3P_J) |{}^3S_1\rangle
+{H_1^1({}^3S_1) \over m^6}
        \langle {}^3S_1| {\cal Q}_1^1({}^3S_1) |{}^3S_1\rangle\nonumber\\
&&\hbox{}+{H_1^2({}^3S_1) \over m^6}
        \langle {}^3S_1| {\cal Q}_1^2({}^3S_1) |{}^3S_1\rangle.
\label{Gpsilh}
\end{eqnarray}
The operator ${\cal O}_8({}^1S_0)$ is defined in Eq.~(\ref{o81s0}), and
the operator ${\cal O}_8({}^3S_1)$ is defined in Eq.~(\ref{o83s1}). The
remaining operators in Eq.~(\ref{Gpsilh}) are defined by
\begin{subequations}
\label{3s1-ops}
\begin{eqnarray}
{\cal O}_1({}^3S_1)&=&\psi^\dagger\bm{\sigma}\chi\cdot
\chi^\dagger\bm{\sigma}\psi,\\
{\cal P}_1({}^3S_1)&=&{1\over 2}\left[\psi^\dagger\bm{\sigma}\chi\cdot
\chi^\dagger \bm{\sigma}(-\ihalf \tensor{\bf D})^2\psi+
\psi^\dagger\bm{\sigma}(-\ihalf \tensor{\bf D})^2\chi\cdot
\chi^\dagger \bm{\sigma}\psi
\right],\\
{\cal O}_8({}^3P_0)&=&{1\over 3}\psi^\dagger(-\ihalf\tensor{\bf D}
\cdot\bm{\sigma})T_a
\chi\chi^\dagger(-\ihalf\tensor{\bf D}\cdot\bm{\sigma})T_a\psi,\\
{\cal O}_8({}^3P_1)&=&{1\over 2}\psi^\dagger(-\ihalf\tensor{\bf D}
\times\bm{\sigma})T_a\chi\cdot
\chi^\dagger(-\ihalf\tensor{\bf D}\times \bm{\sigma})T_a\psi,\\
{\cal O}_8({}^3P_2)&=&\psi^\dagger(-\ihalf\tensor D^{(i}\sigma^{j)})T_a\chi
\chi^\dagger(-\ihalf\tensor D^{(i}\sigma^{j)})T_a \psi,\\
{\cal Q}_1^1({}^3S_1)&=&\psi^\dagger
\bm{\sigma}(-\ihalf\tensor{\bf D})^2\chi\cdot
\chi^\dagger\bm{\sigma}(-\ihalf\tensor{\bf D})^2\psi,\\
{\cal Q}_1^2({}^3S_1)&=&{1\over 2}\left[\psi^\dagger\bm{\sigma}\chi\cdot
\chi^\dagger\bm{\sigma}(-\ihalf\tensor{\bf D})^4\psi
+\psi^\dagger\bm{\sigma}(-\ihalf\tensor{\bf D})^4\chi\cdot
\chi^\dagger\bm{\sigma}\psi\right],\\
{\cal Q}_1^3({}^3S_1)&=&{1\over 2}\left[\psi^\dagger\bm{\sigma}\chi
\chi^\dagger\cdot\bm{\sigma}(\tensor{\bf D}\cdot g{\bf E}+g{\bf E}\cdot 
\tensor{\bf D})\psi
-\psi^\dagger\bm{\sigma}(\tensor{\bf D}\cdot g{\bf E}+g{\bf E}\cdot 
\tensor{\bf D})\chi\cdot
\chi^\dagger\bm{\sigma}\psi\right].\nonumber\\
\end{eqnarray}
\end{subequations}
The operator ${\cal Q}_1^3({}^3S_1)$ does not appear in
Eq.~(\ref{Gpsilh}) because, as we show in Appendix~\ref{app:relation},
it can be eliminated in favor of ${\cal Q}_1^1({}^3S_1)$ and ${\cal
Q}_1^2({}^3S_1)$ through the use of the equations of motion. From the
velocity-scaling rules in Ref.~\cite{bbl}, we find that, in the
${}^3S_1$ state, the operator ${\cal O}_1({}^3S_1)$ has a matrix element
of relative order  $v^0$, the operator ${\cal P}_1({}^3S_1)$ has a
matrix element of relative order  $v^2$, the operator ${\cal
O}_8({}^1S_0)$ has a matrix element of relative order  $v^3$, and the
operators ${\cal O}_8({}^3S_1)$, ${\cal O}_8({}^3P_0)$, ${\cal
O}_8({}^3P_1)$, ${\cal O}_8({}^3P_2)$, ${\cal Q}_1^1({}^3S_1)$, ${\cal
Q}_1^2({}^3S_1)$, and ${\cal Q}_1^3({}^3S_1)$ have matrix elements of
relative order  $v^4$.

The order-$\alpha_s^3$ and order-$\alpha_s^4$ contributions to the 
short-distance coefficient $F_1({}^3S_1)$ were computed 
by Mackenzie and Lepage \cite{Mackenzie:1981sf} and can be found in 
Ref.~\cite{bbl}, as can the order-$\alpha^2$ contribution:
\begin{eqnarray}
F_1({}^3S_1)&=&
{(N_c^2-1)(N_c^2 - 4)\over N_c^3}{(\pi^2-9)\over 18}\alpha_s^3(m)
\bigg\{ 1 +[- 9.46(2) C_F + 4.13(17) C_A\nonumber\\
&&\hbox{}- 1.161(2) n_f] {\alpha_s \over \pi} \bigg\}
+2\pi Q^2 \left( \sum_{i=1}^{n_f} Q_i^2 \right) \alpha^2
\bigg[ 1 - {13 \over 4} C_F {\alpha_s \over \pi} \bigg],
\label{fpsiNLO}
\end{eqnarray}
where $Q$ is the electric charge of the heavy quark, and the $Q_i$ are
the electric charges of the light quarks. The order-$\alpha_s^3$ 
contribution to the short-distance coefficient $G_1({}^3S_1)$ is
computed in Ref.~\cite{Keung:jb}:
\begin{equation}
G_1({}^3S_1)=-{5(19\pi^2-132)\over 729}\alpha_s^3.
\end{equation}
To leading order in $\alpha_s^2$,
$G_1({}^3S_1)/[m^2F_1({}^3S_1)]=-(19\pi^2-132)/[12(\pi^2-9)]\approx
-5.32$. Hence, the relativistic correction to $J/\psi$ decay is greater
in magnitude than the leading contribution. This situation casts some
doubt on the validity of the $v$ expansion. We investigate this issue
further in this paper by calculating corrections of relative order
$v^4$. The order-$\alpha_s^2$ and order-$\alpha_s^3$ contributions to
the short-distance coefficients $F_8({}^1S_0)$ and $F_8({}^3S_1)$ are
given in Eqs.~(\ref{f81s0}) and (\ref{f83s1}), respectively. The
order-$\alpha_s^2$ and order-$\alpha_s^3$ contributions to the
short-distance coefficients $F_8({}^3P_J)$ have been computed by
Petrelli {\it et al.} \cite{Petrelli:1998ge}:
\begin{subequations}
\begin{eqnarray}
F_8({}^3P_0)&=&6B_F\pi \alpha_s^2(\mu)
\bigg\{1+{\alpha_s\over\pi}\bigg[\bigg(-{7\over 3}
+{\pi^2\over 4}\bigg)C_F\nonumber\\
&&\qquad\qquad\qquad\hbox{}+\bigg({463\over 81}+{35\over 27}
\log 2-{17\over 216}\pi^2\bigg)C_A
+2b_0\log{\mu\over 2m}\bigg]\bigg\}\nonumber\\
&&\qquad\qquad\qquad\hbox{}+{8\over 9}n_fB_F\alpha_s^3
\bigg(-{29\over 6}+\log{2m\over \mu_\Lambda}
\bigg),\\
F_8({}^3P_1)&=&C_AB_F\alpha_s^3\bigg({1369\over 54}-{23\over 9}
\pi^2\bigg)
+{8\over 9}n_fB_F\alpha_s^3
\bigg(-{4\over 3}+\log{2m\over \mu_\Lambda}\bigg),\\
F_8({}^3P_2)&=&{8B_F\pi\over 5} \alpha_s^2(\mu)
\bigg\{1+{\alpha_s\over\pi}\bigg[-4C_F+
\bigg({4955\over 431}+{7\over 9}\log 2-{43\over 72}\pi^2\bigg)C_A
\nonumber\\
&&\qquad\qquad\qquad\hbox{}+2b_0\log{\mu\over 2m}\bigg]\bigg\}
+{8\over 9}n_fB_F\alpha_s^3\bigg(-{29\over 15}+\log{2m\over\mu_\Lambda}
\bigg),
\end{eqnarray}
\end{subequations}
where $\mu_\Lambda$ is the NRQCD renormalization scale. The contribution
to $F_8({}^1P_1)$ of order $\alpha_s^2$ vanishes because Yang's theorem
\cite{yang} forbids the decay of a spin-one particle into two equivalent
massless vector particles (gluons). The contributions from decay into a
light quark-antiquark pair vanish because the ${}^3P_J$ states are even
under charge conjugation. Again, the individual quantities
$H_1^1({}^3S_1)$ and $H_1^2({}^3S_1)$ cannot be distinguished in
processes in which the heavy quark and antiquark decay on shell. We
compute the quantity $H_1^1({}^3S_1)+H_1^2({}^3S_1)$ in this paper.

Through relative-order $v^4$, the decay of a ${}^1S_0$ state into two photons 
is given by
\begin{eqnarray}
\Gamma({}^1S_0 \to \gamma \gamma)
&=&{F_{\gamma\gamma}({}^1S_0) \over m^2}
\Big| \langle 0|\chi^\dagger \psi |{}^1S_0\rangle \Big|^2 \nonumber\\
&&\hbox{} +{G_{\gamma\gamma}({}^1S_0) \over m^4} 
{\rm \, Re \,}\left[ \langle{}^1S_0| \psi^\dagger \chi |0\rangle
\langle 0| \chi^\dagger (-\ihalf \tensor{\bf D})^2 \psi 
|{}^1S_0\rangle \right]\nonumber\\
&&\hbox{} +{H_{\gamma\gamma}^1({}^1S_0) \over m^6} 
\langle{}^1S_0| \psi^\dagger(-\ihalf\tensor{\bf D})^2 
\chi |0\rangle
\langle 0| \chi^\dagger (-\ihalf \tensor{\bf D})^2 \psi 
|{}^1S_0\rangle\nonumber\\
&&\hbox{} +{H_{\gamma\gamma}^2({}^1S_0) \over m^6} 
{\rm \, Re \,}\left[ \langle{}^1S_0| \psi^\dagger \chi |0\rangle
\langle 0| \chi^\dagger (-\ihalf \tensor{\bf D})^4 \psi 
|{}^1S_0\rangle \right].
\label{Getagg}
\end{eqnarray}
The product of matrix elements ${\rm \, Re \,}\left[ \langle{}^1S_0|
\psi^\dagger \chi |0\rangle \langle 0| \chi^\dagger (\tensor{\bf D}\cdot
g{\bf E}+g{\bf E}\cdot \tensor{\bf D}) \psi |{}^1S_0\rangle \right]$,
which is of relative order $v^4$, does not appear in Eq.~(\ref{Getagg})
because, as we show in Appendix~\ref{app:relation}, it can be
eliminated in favor of the products of matrix elements in the last two 
terms of Eq.~(\ref{Getagg}) through the use of the equations of
motion. From the velocity-scaling rules in Ref.~\cite{bbl}, we find
that, in Eq.~(\ref{Getagg}), the product of matrix elements in the first
line is of relative order $v^0$, the product of matrix elements in the
second line is of relative order $v^2$, and the products of matrix
elements in the third and fourth lines are of relative order $v^4$.

The order-$\alpha^2$ and order-$\alpha^2\alpha_s$ contributions to the
short-distance coefficient $F_{\gamma\gamma}({}^1S_0)$ are calculated in
Refs.~\cite{harris-brown,Barbieri:1979be,Hagiwara:1980nv} and are given
in Ref.~\cite{bbl}:
\begin{equation}
F_{\gamma\gamma}({}^1S_0)=2\pi Q^4\alpha^2\left[1+\left({\pi^2\over 
4}-5\right)C_F{\alpha_s\over\pi}\right].
\end{equation}
The order-$\alpha^2$ contribution to $G_{\gamma\gamma}({}^1S_0)$ is 
computed in Refs.~\cite{Keung:jb,bbl}:
\begin{equation}
G_{\gamma\gamma}({}^1S_0)=-{8\pi Q^4\over 3}\alpha^2.
\end{equation}
To leading order in $\alpha_s$, $G_{\gamma\gamma}({}^1S_0)/[m^2
F_{\gamma\gamma}({}^1S_0)]=-4/3$. Hence, the first relativistic
correction to this process is substantial for the $\eta_c$. In this
paper, we compute the combination of short-distance coefficients
$H_{\gamma\gamma}^1({}^1S_0)+H_{\gamma\gamma}^2({}^1S_0)$.

Through relative order $v^4$, the rate for a ${}^3S_1$ state to decay
into an $e^+e^-$ pair is
\begin{eqnarray}
\Gamma({}^3S_1 \to e^+ e^-)
&=& {F_{ee}({}^3S_1) \over m^2} 
        \Big| \langle 0| \chi^\dagger \bm{\sigma} \psi |{}^3S_1\rangle 
\Big|^2 \nonumber\\
&&\hbox{}+{G_{ee}({}^3S_1) \over m^4}
        {\rm \, Re \,}\left[ \langle{}^3S_1| \psi^\dagger \bm{\sigma} \chi 
|0\rangle \cdot
        \langle 0| \chi^\dagger \bm{\sigma} (-\ihalf 
\tensor{\bf D})^2 
\psi |{}^3S_1\rangle \right]\nonumber\\
&&\hbox{}+{H_{ee}^1({}^3S_1) \over m^6}
\langle{}^3S_1| \psi^\dagger \bm{\sigma}
(-\ihalf\tensor{\bf D})^2 \chi 
|0\rangle \cdot
        \langle 0| \chi^\dagger \bm{\sigma} (-\ihalf\tensor{\bf D})^2 
\psi |{}^3S_1\rangle\nonumber\\
&&\hbox{}+{H_{ee}^2({}^3S_1) \over m^6}
        {\rm \, Re \,}\left[ \langle{}^3S_1| \psi^\dagger \bm{\sigma} \chi 
|0\rangle \cdot
        \langle 0| \chi^\dagger \bm{\sigma} (-\ihalf \tensor{\bf D})^4 
\psi |{}^3S_1\rangle \right].
\label{Gpsiee}
\end{eqnarray}

The product of matrix elements ${\rm \, Re \,}\left[ \langle{}^3S_1|
\psi^\dagger \bm{\sigma}\chi|0\rangle \cdot \langle 0| \chi^\dagger
\bm{\sigma}( \tensor{\bf D}\cdot g{\bf E}+g{\bf E}\cdot \tensor {\bf D})
\psi |{}^3S_1\rangle \right]$, which is of relative order $v^4$, does
not appear in Eq.~(\ref{Gpsiee}) because, as we show in
Appendix~\ref{app:relation}, it can be eliminated in favor of the
products of matrix element in the last two terms of Eq.~(\ref{Gpsiee})
through the use of the equations of motion. In Eq.~(\ref{Gpsiee}), the
product of matrix elements in the first line is of relative order $v^0$,
the product of matrix elements in the second line is of relative order
$v^2$, and the products of matrix elements in the third and fourth lines
are of relative order $v^4$.

The order-$\alpha^2$ and order-$\alpha^2\alpha_s$ contributions to the
short-distance coefficient $F_{ee}({}^3S_1)$ are calculated in
Refs.~\cite{Barbieri:1975ki,Celmaster:1978yz} and are given in
Ref.~\cite{bbl}. The order-$\alpha^2\alpha_s^2$ contribution is
calculated in Ref.~\cite{Beneke:1997jm}. Altogether, these contributions
give
\begin{equation}
F_{ee}({}^3S_1)={2\pi Q^2\alpha^2\over 3}\left\{1-4C_F
{\alpha_s(m)\over\pi}
+\left[-117.46+0.82n_f+{140\pi^2\over 27}
\ln\left({2m\over\mu_\Lambda}\right)\right]
\left({\alpha_s\over\pi}\right)^2\right\}.
\end{equation}
The order-$\alpha^2$ contribution to $G_{\gamma\gamma}({}^1S_0)$ is 
computed in Refs.~\cite{Keung:jb,bbl}:
\begin{equation}
G_{ee}({}^3S_1)=-{8\pi Q^2\over 9}\alpha^2.
\end{equation}
To leading order in $\alpha_s$, $G_{ee}({}^3S_1)/[m^2 F_{\rm
ee}({}^3S_1)]=-4/3$. Hence, the first relativistic correction to this
process is substantial for the $J/\psi$. In this paper, we compute the
combination of short-distance coefficients
$H_{ee}^1({}^3S_1)+H_{ee}^1({}^3S_1)$.

\section{Spin projectors}

In computing the quarkonium decay rates, we use the covariant
spin-projector method \cite{Kuhn:1979bb,Guberina:1980dc} to identify
spin-singlet and spin-triplet amplitudes. For purposes of the
computations in this paper, we need projection operators accurate at
least through relative order $v^4$. In this section, we compute the
required projectors to all orders in $v$.

The Dirac spinors, with the standard nonrelativistic normalization, may 
be written as 
\begin{subequations}
\label{spinors}
\begin{eqnarray}
u({\bf p}) &=& \sqrt{E+m \over 2E}
\left( \begin{array}{c} \xi \\
        {{\bf p} \cdot \bm{\sigma} \over E+m} \xi \end{array} \right) ,
\label{uspinor}\\
v(-{\bf p}) &=& \sqrt{E+m \over 2E}
\left( \begin{array}{c} {(-{\bf p}) \cdot \bm{\sigma} \over E+m} \eta \\ 
\eta  \end{array} \right) ,
\label{vspinor}
\end{eqnarray}
\end{subequations}
where $\xi$ and $\eta$ are two-component Pauli spinors, and 
$E(p)=\sqrt{m^2+{\bf p}^2}$. 
We take the heavy quark and antiquark momenta to be
\begin{subequations}
\label{momenta}
\begin{eqnarray}
p_Q&=&(1/2)P+p,\\
p_{\bar Q}&=&(1/2)P-p,
\end{eqnarray}
\end{subequations}
respectively, where in the quarkonium rest frame,
\begin{subequations}
\label{rest-frame}
\begin{eqnarray}
P&=&(2E(p),\bm{0}),\\
p&=&(0,{\bf p}).
\end{eqnarray}
\end{subequations}

Using Eq.~(\ref{spinors}), it is straightforward to express spin-singlet
and spin-triplet combinations of spinor bilinears in terms of Dirac
matrices and to write them in a covariant form. In the spin-singlet
case, we have
\begin{eqnarray}
\Pi_0(P,p)&=&-\sum_{\lambda_1,\lambda_2}u({\bf p},\lambda_1)
\bar v(-{\bf p},\lambda_2)
\langle \half\,\lambda_1\,\half\,\lambda_2|0\,0\rangle\nonumber\\
&=&{1\over \sqrt 2}{E+m\over 2E}
\left(1+{\bm{\alpha}\cdot{\bf p}\over E+m}\right)
{1+\gamma_0\over 2}\gamma_5
\left(1-{\bm{\alpha}\cdot{\bf p}\over E+m}\right)\gamma_0\nonumber\\
&=&{1\over 2\sqrt{2}E(E+m)}\left(\half\not\!P+m+\not\!p\right)
{\not\!P+2E\over 4E}\gamma_5
\left(\half\not\!P-m-\not\!p\right),
\label{singlet-projector}
\end{eqnarray}
where $\alpha_i$ and $\gamma_\mu$ are Dirac matrices in the Dirac
representation, $\gamma_5=i\gamma^0\gamma^1\gamma^2\gamma^3$, and we
have chosen the normalization so that the projector
(\ref{singlet-projector}) corresponds in NRQCD to the projector $I/\sqrt
2$, where $I$ is a unit Pauli matrix.\footnote{In
Eq.~(\ref{singlet-projector}), the standard Clebsch-Gordan coefficients
are appropriate if the spinors in Eq.~(\ref{spinors}) are related to
each other through a unitary transformation, which preserves the SU(2)
algebra, such as the charge-conjugation transformation
$\eta=-i\sigma_2\xi$. One such choice of spinors is
$\xi=\left(\begin{array}{c}1\\0\end{array}\right)\hbox{ or }
\left(\begin{array}{c}0\\1\end{array}\right)$ and
$\eta=\left(\begin{array}{c}0\\1\end{array}\right)\hbox{ or }
\left(\begin{array}{c}-1\\0\end{array}\right)$, for $\lambda_1=\pm 1/2$
and $\lambda_2=\pm 1/2$, respectively. On the other hand, a popular
convention is $\xi=\left(\begin{array}{c}1\\0\end{array}\right)\hbox{ or
} \left(\begin{array}{c}0\\1\end{array}\right)$ and
$\eta=\left(\begin{array}{c}0\\1\end{array}\right)\hbox{ or }
\left(\begin{array}{c}1\\0\end{array}\right)$, for $\lambda_1=\pm 1/2$
and $\lambda_2=\pm 1/2$, respectively. With this convention, the
Clebsch-Gordan coefficients in Eq.~(\ref{singlet-projector}) must be
multiplied by an additional factor $(-1)^{(-1/2+\lambda_2)}$.} We note
that $E(p)$ may be written in a Lorentz invariant fashion as
\begin{equation}
E(p)=(1/2)\sqrt{P^2}.
\end{equation}
In the case of a spin-triplet state with polarization $\epsilon$, we
have
\begin{eqnarray}
\Pi_1(P,p,\bm{\epsilon})&=&\sum_{\lambda_1,\lambda_2}
u({\bf p},\lambda_1)\bar v(-{\bf p},\lambda_2)
\langle \half\,\lambda_1\,\half\, \lambda_2
|1\,\epsilon\rangle\nonumber\\
&=&{1\over \sqrt 2}{E+m\over 2E}
\left(1+{\bm{\alpha}\cdot{\bf p}\over E+m}\right)
{1+\gamma_0\over 2}\bm{\alpha}\cdot \bm{\epsilon}
\left(1-{\bm{\alpha}\cdot{\bf p}\over E+m}\right)\gamma_0\nonumber\\
&=&{-1\over 2\sqrt{2}E(E+m)}\left(\half\not\!P+m+\not\!p\right)
{\not\!P+2E\over 4E}\not\!\epsilon
\left(\half\not\!P-m-\not\!p\right).
\label{triplet-projector}
\end{eqnarray}
Here, $|1\,\epsilon\rangle$ is the rotationally invariant linear
combination
$|1\,\epsilon\rangle=\epsilon^-|1\,1\rangle-\epsilon^+|1\,-1\rangle
-\epsilon_3|1\,0\rangle$, with $\epsilon^\pm=(1/\sqrt 2) (\epsilon_1\pm
i\epsilon_2)$. We have chosen the normalization so that the projector
(\ref{triplet-projector}) corresponds in NRQCD to the projector
$\bm{\sigma}\cdot \bm{\epsilon}/\sqrt{2}$. The expressions
(\ref{singlet-projector}) and (\ref{triplet-projector}) are valid to all
orders in $v$.

\section{Relativistic corrections to ${}^1S_0$ decays}

In this section we compute the short-distance coefficients that
appear in the corrections through relative order $v^4$ to ${}^1S_0$
quarkonium decays into two photons and into light hadrons (two gluons).

We begin with the case of decay into two photons. We take the
definitions of the heavy quark and antiquark momenta given in
Eq.~(\ref{momenta}) and work in the quarkonium rest frame, as defined in
Eq.~(\ref{rest-frame}). We take the outgoing photon momenta to be $k$ 
and $q$, with polarization indices $\mu$ and $\nu$, respectively. 
Consider first the diagram in which the quark emits the photon with 
momentum $k$. The spin-singlet amplitude corresponding to this diagram is
\begin{eqnarray}
A_1({\rm sing}\rightarrow \gamma\gamma)&=&-ie^2Q^2{\rm Tr}
\left[\Pi_0(P,p)\gamma^\nu {\not\!p_Q-\not\!k+m\over -2p_Q\cdot k}
\gamma^\mu\right]\nonumber\\
&=&{-ie^2Q^2\over 2\sqrt{2}E(E+m)}{1\over 2p_Q\cdot k}{\rm Tr}
\left[\gamma^\nu \not\!k \gamma^\mu(\half\not\!P+m+\not\!p)
{1+\gamma_0\over 2}\gamma_5(\half\not\!P-m-\not\!p)\right]\nonumber\\
&=&{-ie^2Q^2\over 2\sqrt{2}E(E+m)}{1\over 2p_Q\cdot k}{\rm Tr}
\left[\gamma^\nu \not\!k \gamma^\mu(E+m+\not\!p){1+\gamma_0\over
2}\gamma_5(-E-m-\not\!p)\right],\nonumber\\
\end{eqnarray}
where $e$ is the electromagnetic coupling constant.\footnote{In 
computing the short-distance coefficients for the electromagnetic decay 
processes, we suppress trival color factors, which ultimately cancel 
when one matches decay rates in full QCD and NRQCD.}
In the projector $(1+\gamma_0)/2$ in the last line, the term 
proportional to $1$ gives a vanishing trace, while the term proportional 
to $\gamma_0$ gives
\begin{equation}
A_1({\rm sing}\rightarrow \gamma\gamma)=-e^2Q^2{m\over \sqrt{2}E}
\epsilon^{\nu\rho\mu 0}k_\rho {1\over p_Q\cdot k}.
\end{equation}
Similarly, the diagram in which the antiquark emits the photon with 
momentum $k$ yields an amplitude
\begin{equation}
A_2({\rm sing}\rightarrow \gamma\gamma)=e^2Q^2{m\over \sqrt{2}E}
\epsilon^{\mu\rho\nu 0}k_\rho {1\over p_{\bar Q}\cdot k}.
\end{equation}
Adding $A_1({\rm sing}\rightarrow \gamma\gamma)$ and 
$A_2({\rm sing}\rightarrow \gamma\gamma)$, we obtain the complete 
amplitude for ${}^1S_0$ charmonium decay into two photons:
\begin{equation}
A({\rm sing}\rightarrow \gamma\gamma)=-e^2Q^2{m\over \sqrt{2}E}
\epsilon^{\nu\rho\mu 0}k_\rho \left({1\over E^2-{\bf p}\cdot {\bf k}}
+{1\over E^2+{\bf p}\cdot {\bf k}}\right).
\end{equation}
We project out the $S$-wave part of the amplitude by averaging over the 
angles of ${\bf p}$:
\begin{eqnarray}
A({}^1S_0 \rightarrow \gamma\gamma)&=&{1\over 2}\int_{-1}^1
d(\cos\theta)(-e^2Q^2){m\over \sqrt{2}E}
\epsilon^{\nu\rho\mu 0}k_\rho \left({1\over E^2-|{\bf p}|
|{\bf k}|\cos\theta}+{1\over E^2+|{\bf p}||{\bf k}|\cos\theta}
\right)\nonumber\\
&=&-e^2Q^2\epsilon^{\nu\rho\mu 0}k_\rho {m\over \sqrt{2}E^2 |{\bf p}|}
\ln{E+|{\bf p}|\over E-|{\bf p}|},
\label{A-1s0-pp}
\end{eqnarray}
where we have used ${\bf k}^2=E^2$. Multiplying the expression
(\ref{A-1s0-pp}) by its complex conjugate, by the two-body phase space
$1/(8\pi)$, and by a factor $1/2!$ for two identical particles in the
final state, we obtain the decay width for a ${}^1S_0$ $Q\bar Q$ state
into two photons:
\begin{equation}
\Gamma({}^1S_0 \rightarrow \gamma\gamma)={\pi m^2Q^4\alpha^2\over
E^2 {\bf p}^2}\ln^2{E+|{\bf p}|\over E-|{\bf p}|}.
\label{gamma-1s0-pp}
\end{equation}
Here, and in succeeding computations of the decay widths of two-particle
states, we suppress a factor of the inverse volume that is associated
with the normalization of the initial state.

From Eq.~(\ref{Getagg}), we find that the decay width for a ${}^1S_0$
$Q\bar Q$ state into two photons in NRQCD in order $\alpha_s^0$ and
through relative order $v^4$ is
\begin{eqnarray}
\Gamma_{\rm NRQCD}({}^1S_0 \rightarrow \gamma\gamma)&=&
2[(1/m^2)F_{\gamma\gamma}({}^1S_0)
+({\bf p}^2/m^4) G_{\gamma\gamma}({}^1S_0)
+({\bf p}^4/m^6) H^1_{\gamma\gamma}({}^1S_0)\nonumber\\
&&\qquad\hbox{}+({\bf p}^4/m^6) H^2_{\gamma\gamma}({}^1S_0)],
\label{gamma-nrqcd-1s0-pp}
\end{eqnarray}
where the factor two on the right side of Eq.~(\ref{gamma-nrqcd-1s0-pp})
comes from the spin factor for normalized heavy-quark states.

Comparing powers of ${\bf p}^2/m^2$ in Eqs.~(\ref{gamma-1s0-pp}) and 
(\ref{gamma-nrqcd-1s0-pp}), we obtain the short-distance coefficients at
leading order in $\alpha_s$:
\begin{subequations}
\label{1s0-pp-coeffs}
\begin{eqnarray}
F_{\gamma\gamma}({}^1S_0)&=&2\pi Q^4\alpha^2,\\
G_{\gamma\gamma}({}^1S_0)&=&-{8\pi\over 3} Q^4 \alpha^2,\\
H^1_{\gamma\gamma}({}^1S_0)+H^2_{\gamma\gamma}({}^1S_0)&=&
{136\pi\over 45}Q^4 \alpha^2.
\end{eqnarray}
\end{subequations}
Our results for $F_{\gamma\gamma}({}^1S_0)$ and 
$G_{\gamma\gamma}({}^1S_0)$ confirm those given in 
Refs.~\cite{bbl,harris-brown,Barbieri:1979be,Hagiwara:1980nv} and 
\cite{bbl}, respectively. Our result for 
$H^1_{\gamma\gamma}({}^1S_0)+H^2_{\gamma\gamma}({}^1S_0)$ is new.

At leading order in $\alpha_s$, the decay of a ${}^1S_0$ $Q\bar Q$ state
to light hadrons proceeds through an annihilation into two gluons.
Hence, we may obtain the decay width for a ${}^1S_0$ $Q\bar Q$ state
into light hadrons by multiplying the width into two photons
[Eq.~(\ref{gamma-1s0-pp})] by a color factor $C_F/2$ times
$\alpha_s^2/(\alpha^2 Q^4)$:
\begin{equation}
\Gamma({}^1S_0 \rightarrow {\rm LH})={\pi C_F m^2\alpha_s^2\over
2E^2 {\bf p}^2}\ln^2{E+|{\bf p}|\over E-|{\bf p}|}.
\label{gamma-1s0-lh}
\end{equation}
From Eq.~(\ref{eta-decay}), we find that the decay width for a ${}^1S_0$
$Q\bar Q$ state into two photons in NRQCD in order $\alpha_s^2$ and
through relative order $v^4$ is
\begin{eqnarray}
\Gamma_{\rm NRQCD}({}^1S_0 \rightarrow {\rm LH})&=&
2N_c[(1/m^2)F_1({}^1S_0)
+({\bf p}^2/m^4) G_1({}^1S_0)
+({\bf p}^4/m^6) H^1_1({}^1S_0)\nonumber\\
&&\qquad\hbox{}+({\bf p}^4/m^6) H^2_1({}^1S_0)],
\label{gamma-nrqcd-1s0-lh}
\end{eqnarray}
where the factor $2N_c$ on the right side of
Eq.~(\ref{gamma-nrqcd-1s0-lh}) comes from the spin and color factors for
normalized heavy-quark states. The matrix elements of the color-octet
operators do not contribute to Eq.~(\ref{gamma-nrqcd-1s0-lh}) in order
$\alpha_s^2$. Comparing Eqs.~(\ref{gamma-1s0-lh}) and
(\ref{gamma-nrqcd-1s0-lh}), we obtain the short-distance coefficients at
leading order in $\alpha_s$:
\begin{subequations}
\label{1s0-gg-coeffs}
\begin{eqnarray}
F_1({}^1S_0)&=&{\pi C_F\over N_c}\alpha_s^2,\\
G_1({}^1S_0)&=&-{4\pi C_F\over 3N_c} \alpha_s^2,\\
H_1^1({}^1S_0)+H_1^2({}^1S_0)&=&{68\pi C_F\over 45N_c}\alpha_s^2.
\end{eqnarray}
\end{subequations}
Our result for $F_1({}^1S_0)$ is in agreement with that given in
Refs.~\cite{Barbieri:1979be,Hagiwara:1980nv,bbl}, and our
result for $G_1({}^1S_0)$ is in agreement with that given in
Ref.~\cite{bbl}. Our result for $H_1^1({}^1S_0)+H_1^2({}^1S_0)$ is new.

\section{Relativistic corrections to ${}^3S_1$ decay to $e^+e^-$}

Next we turn to the case of the decay of a ${}^3S_1$ quarkonium state
into an $e^+e^-$ pair. Again, we work in the quarkonium rest frame
defined in Eq.~(\ref{rest-frame}). The amplitude for a quark and
antiquark in a spin-triplet state with the momenta given in
Eq.~(\ref{momenta}) to decay into a virtual photon with polarization
index $\mu$ is given by
\begin{eqnarray}
A({\rm trip}\rightarrow \gamma^*)&=&ieQ\,{\rm Tr}\,[\Pi_1(P,p,\epsilon)
\gamma_\mu]\nonumber\\
&=&ieQ\sqrt{2}\left[{p_\mu p\cdot\epsilon\over E(E+m)}
+\epsilon_\mu\right].
\end{eqnarray}
We can project out the $S$-wave part of the amplitude by averaging over 
the angles of ${\bf p}$:
\begin{eqnarray}
A({}^3S_1\rightarrow \gamma^*)&=&{1\over 2}\int_{-1}^1 d(\cos\theta)\,
ieQ\sqrt{2}\left[{p_\mu p\cdot\epsilon\over E(E+m)}
+\epsilon_\mu\right]\nonumber\\
&=& ieQ\sqrt{2}\left({2\over 3}+{m\over 3E}\right)\epsilon_\mu.
\label{A-3s1-p}
\end{eqnarray}
In order to obtain the decay rate into an $e^+e^-$ pair, we multiply the
expression (\ref{A-3s1-p}) by its complex conjugate with index $\nu$, by
a photon-propagator factor $-ig_{\mu\rho}/k^2$, by a complex-conjugated
photon-propagator factor $ig_{\nu\sigma}/k^2$, and by twice the
imaginary part of the $e^+e^-$-pair contribution to the photon's vacuum
polarization, namely, $(g_{\rho\sigma}k^2-k_\rho k_\sigma)(-2/3)\alpha$.
Here $k$ is the virtual photon's momentum. The result is
\begin{equation}
\Gamma({}^3S_1\rightarrow e^+e^-)=
{4\pi Q^2\alpha^2\over 3E^2}\left({2\over 3}+{m\over 3E}\right)^2,
\label{gamma-3s1-ee}
\end{equation}
where we have used $k\cdot\epsilon=0$, $\epsilon\cdot\epsilon^*=-1$, and
$k^2=4E^2$.

From Eq.~(\ref{Gpsiee}), we see that, in NRQCD through relative order 
$v^4$, the decay width for a ${}^3S_1$ $Q\bar Q$ state into an $e^+e^-$ 
pair is
\begin{eqnarray}
\Gamma_{\rm NRQCD}({}^3S_1 \rightarrow e^+e^-)&=&
2 [(1/m^2)F_{ee}({}^3S_1)
+({\bf p}^2/m^4) G_{ee}({}^3S_1)
+({\bf p}^4/m^6) H^1_{ee}({}^3S_1)\nonumber\\
&&\qquad\hbox{}+({\bf p}^4/m^6) H^2_{ee}({}^3S_1)].
\label{gamma-nrqcd-3s1-ee}
\end{eqnarray}
The factor two on the right side of Eq.~(\ref{gamma-nrqcd-3s1-ee}) comes 
from the spin factor for normalized heavy-quark states. 

Comparing powers of ${\bf p}^2/m^2$ in 
Eqs.~(\ref{gamma-3s1-ee}) and (\ref{gamma-nrqcd-3s1-ee}), 
we obtain the short-distance coefficients at leading order in 
$\alpha_s$:
\begin{subequations}
\label{3s1-ee-coeffs}
\begin{eqnarray}
F_{ee}({}^3S_1)&=&{2\pi\over 3}Q^2\alpha^2,\\
G_{ee}({}^3S_1)&=&-{8\pi\over 9}Q^2\alpha^2,\\
H^1_{ee}({}^3S_1)+H^2_{ee}({}^3S_1)&=&
{58\pi\over 54}Q^2\alpha^2.
\end{eqnarray}
\end{subequations}
Our result for $F_{ee}({}^3S_1)$ agrees with that given in
Refs.~\cite{bbl,Barbieri:1975ki,Celmaster:1978yz}, and our result for
$G_{ee}({}^3S_1)$ agrees with that given in Refs.~\cite{bbl}. Our result
for $H^1_{ee}({}^3S_1)+H^2_{ee}({}^3S_1)$ is new.

\section{Relativistic corrections to ${}^3S_1$ decay to light hadrons}

In the decay of a heavy-quark-antiquark state, diagrams in which only
two of the final-state gluons attach to the heavy-quark line have a
common heavy-quark color factor. Hence, (Abelian) charge-conjugation
symmetry forbids such diagrams in the decay of a ${}^3S_1$ state.
Furthermore, color conservation forbids diagrams in which only one of
the final-state gluons attaches to the heavy-quark line. Thus, in
leading order in $\alpha_s$, a ${}^3S_1$ heavy-quark-antiquark state
decays into three gluons, and the decay proceeds through diagrams
in which all three gluons attach to the heavy-quark line. (Since no 
triple-gluon vertices appear, there are no ghost contributions.)

In this decay process, in contrast with the decay processes that we have
analyzed in the preceding sections, the kinematics allow one of the
final-state gluons to have zero energy. Hence, the possibility arises
that the decay rate contains an infrared (IR) divergence. Simple power 
counting arguments show that an IR divergence can arise only if the soft 
gluon attaches to an incoming (on-shell) heavy-quark or heavy-antiquark leg. 
Therefore, one can use NRQCD to analyze the interaction of this soft gluon 
with the heavy quark. 

One can see from power-counting arguments, that, through relative order
$v^4$, in the Coulomb gauge, a gluon that interacts with a quark or an
antiquark can yield an IR divergence only if the interactions are of the
type $\psi^\dagger {\bf D}\cdot {\bf A}\psi$ or $\chi^\dagger {\bf
D}\cdot {\bf A}\chi$. (The $\psi^\dagger {\bf B}\cdot \bm{\sigma}\psi$
and $\chi^\dagger {\bf B}\cdot \bm{\sigma}\chi$ interactions have the
correct dimensions to produce an IR divergence, but the factors of ${\bf
B}$ bring in powers of the gluon momentum that protect against an IR
divergence.) The factor of ${\bf D}$ translates into a factor of the
incoming quark or antiquark momentum. Factors of the gluon momentum do
not appear since they are orthogonal to the gluon propagator in the
Coulomb gauge. Therefore, the interactions of the gluon yield two
factors of the incoming quark or antiquark momentum in the squared
amplitude. Two additional factors of the incoming quark or antiquark
momentum are required in order to have a nonzero overlap with an
incoming $S$-wave state. Hence, an IR divergence in the decay rate must
be associated with at least four factors of the incoming quark or
antiquark momentum. That is, an IR divergence can first appear in
relative order $v^4$. Because the soft gluon in a ${\bf D}\cdot {\bf A}$
interaction changes the incoming $S$-wave color-singlet quark-antiquark
state into a $P$-wave color-octet quark-antiquark state, we expect that
the IR divergence will be absorbed into matrix elements of the $P$-wave
color-octet operators in Eq.~(\ref{Gpsilh}).

Now let us turn to the actual computation of the rate for a ${}^3S_1$
$Q\bar Q$ state to decay into three gluons. We present only the outlines
of that calculation here.  We used the symbolic manipulation program
\textsc{mathematica} and the package \textsc{feyncalc}
\cite{Mertig:an} to handle the tedious, but straightforward, details of
the algebra.

We regulate the anticipated IR divergence by computing in
$D=4-2\epsilon$ dimensions. We work in the quarkonium rest frame, assign
the incoming quark and antiquark momenta as in Eqs.~(\ref{momenta}) and
(\ref{rest-frame}), and take the outgoing gluon momenta to be $k_1$,
$k_2$, and $k_3$. First we compute the sum of the six Feynman diagrams
for this process, making use of the projector (\ref{triplet-projector}).
Although we are working in $D=4-2\epsilon$ dimensions, we can follow the
approach of Ref.~\cite{Petrelli:1998ge} and simply use the
$D$-dimensional version of the spin-1 projector
(\ref{triplet-projector}). As explained in Ref.~\cite{Petrelli:1998ge},
we need not consider projectors for the higher-spin evanescent NRQCD
operators that appear in $D$ dimensions because the contributions that
contain an IR pole in one loop do not mix the higher-spin operators with
the spin-1 operators.

At this point, we could square the amplitude, integrate over the phase
space, and expand in powers of ${\bf p}$ in order to obtain the desired
result. However, the amount of algebra would be greatly reduced if we
could make the expansion in powers of ${\bf p}$ before carrying out the
phase-space integration. Such a strategy is complicated by the fact that
the phase space depends on ${\bf p}$ through the total energy of the
incoming $Q\bar Q$ state, but we can make that dependence explicit by
introducing a rescaling of phase-space integration variables:
\begin{equation}
k_i\;\rightarrow\; k_i E(p)/m.
\label{rescaling}
\end{equation}
Then, the final-state phase space transforms as
\begin{equation}
\prod_{i}\left({d^{(D-1)}k_i\over 2(k_i)_0}\right)\delta^D (P-\sum_i k_i)
\;\rightarrow\;
\prod_{i}\left({d^{(D-1)}k_i\over 2(k_i)_0}\right)
\delta^D \left({mP\over E(p)}-\sum_i k_i\right)
f(p),
\label{rescaled-ps}
\end{equation}
where
\begin{equation}
f(p)=\left[{E(p)\over m}\right]^{(D-2)^3/D}=\left[{E(p)\over m}\right]^2
\left[1-{5\over 2}\epsilon\log{E^2(p)\over m^2}\right]+O(\epsilon^2).
\label{factor}
\end{equation}
All of the dependence on ${\bf p}$ on the right side of
Eq.~(\ref{rescaled-ps}) is contained in the explicit factor $f(p)$. The
remaining factors correspond to the phase space evaluated at the $Q\bar
Q$ threshold $p=0$. Therefore, after rescaling the $k_i$ according to
Eq.~(\ref{rescaling}), we can obtain the necessary expansion in powers
of ${\bf p}$ by expanding the amplitude, its complex conjugate, and
$f(p)$ in powers of ${\bf p}$ {\it before} carrying out the phase-space
integration. Note that an IR pole in $\epsilon$ first appears in the
rate, excluding the factor $f(p)$, only in the relative-order $v^4$.
Hence, we can drop the term proportional to $\epsilon$ on the right side
of Eq.~(\ref{factor}), which contributes an additional factor $v^2$.

We expand the amplitude in a power series in ${\bf p}$, through order
${\bf p}^4$. The terms containing no powers of ${\bf p}$ yield a pure
$S$-wave contribution. For the terms containing two powers of $p$, we
extract the $S$-wave contribution by making the replacement
\begin{equation}
p_\mu p_\nu\rightarrow {\bf p}^2T_{\mu\nu}.
\end{equation}
For the terms containing four powers of $p$, we extract the $S$-wave
contribution by making the replacement
\begin{equation}
p_\mu p_\nu p_\rho p_\sigma\rightarrow {\bf p}^4T_{\mu\nu\rho\sigma}.
\end{equation}
Here,
\begin{equation}
T_{\mu\nu}={1\over D-1}\Pi_{\mu\nu},
\end{equation}
\begin{equation}
T_{\mu\nu\rho\sigma}={1\over (D-1)(D+1)}
\left[\Pi_{\mu\nu}\Pi_{\rho\sigma}+\Pi_{\mu\rho}\Pi_{\nu\sigma}
+\Pi_{\mu\sigma}\Pi_{\nu\rho}\right],
\end{equation}
and
\begin{equation}
\Pi_{\mu\nu}=-g_{\mu\nu}+{P_\mu P_\nu\over 4E^2(p)}.
\end{equation}
Next, we multiply the amplitude by its complex conjugate. We evaluate
the gluon polarization sums using the Feynman-gauge expression
\begin{equation}
\epsilon_\mu \epsilon^*_\nu=-g_{\mu\nu},
\end{equation}
we evaluate the spin-triplet-state polarization sum using
\begin{equation}
\epsilon_\mu \epsilon^*_\nu=\Pi_{\mu\nu},
\end{equation}
and we divide by $D-1$ to obtain the average over the spin-triplet-state
polarizations. Owing to the charge-conjugation invariance of the
amplitude, only the part of the color factor that is symmetric in the
color indices survives. It is given by
\begin{equation}
{1\over 16N_c}d^{abc}d^{abc}={(N_c^2-1)(N_c^2-4)\over 16N_c^2}.
\end{equation}
Multiplying by this color factor and by $f(p)$, we obtain the 
``squared matrix element''
that must be integrated over the $p=0$ three-body phase space to obtain
the decay rate. We write the coefficients of ${\bf p}^0$, ${\bf p}^2$,
and ${\bf p}^4$ in terms of the invariants
\begin{subequations}
\label{stu}
\begin{eqnarray}
s&=&2k_1\cdot k_2,\\
t&=&2k_1\cdot k_3,\\
u&=&2k_2\cdot k_3,
\end{eqnarray}
\end{subequations}
where, since we have set $p=0$ in these coefficients, the energy-momentum
conservation relation now reads
\begin{equation}
k_1+k_2+k_3=2m.
\end{equation}
The expressions for these coefficients in four dimensions are given in
Appendix~\ref{app:sq-me}.

We re-write the coefficients of ${\bf p}^0$, ${\bf p}^2$, and ${\bf p}^4$ 
in terms of the invariants
\begin{equation}
x_i={2P\cdot k_i\over (2m)^2}\Biggl\vert_{p=0}.
\end{equation}
It follows that
\begin{subequations}
\label{stux}
\begin{eqnarray}
s&=&2m^2(x_1+x_2-x_3),\\
t&=&2m^2(x_1-x_2+x_3),\\
u&=&2m^2(-x_1+x_2+x_3).
\end{eqnarray}
\end{subequations}
The $D$-dimensional three-body phase space for decay of a particle of 
mass $M$ is \cite{Petrelli:1998ge} 
\begin{equation}
d\Phi_{(3)}={M^2\over 2(4\pi)^3}
\left({4\pi\over M^2}\right)^{2\epsilon}{1\over \Gamma(2-2\epsilon)}
\prod_{i=1}^3(1-x_i)^{-\epsilon}dx_i\delta(2-\sum_{i=1}^3 x_i).
\end{equation}
The phase space at $p=0$ is obtained by making the identification $M=2m$.

It is convenient to make a further change of variables, so that the 
limits of integration are independent of the integration variables. To this 
end, we write
\begin{subequations}
\begin{eqnarray}
x_1=x,\\
x_2=1-xy,\\
x_3=1-x(1-y).
\end{eqnarray}
\end{subequations}
This change of variables is particularly useful in analyzing the
infrared singularities, since it avoids the difficulty that, at the
singular points $x_i=0$, the range of integration in one of the
variables $x_j$ ($i\neq j$) vanishes. Now the phase space is
\begin{equation}
d\Phi_{(3)}={M^2\over 2(4\pi)^3}
\left({4\pi\over M^2}\right)^{2\epsilon}{1\over \Gamma(2-2\epsilon)}
[x^2(1-x)y(1-y)]^{-\epsilon}x\,dx\,dy,
\label{xy-ps}
\end{equation}
where $x$ and $y$ range from $0$ to $1$.

In the cases of the terms proportional to ${\bf p}^0$ and ${\bf p}^2$, the 
integrations over the phase space are IR finite, and we can carry out the 
integrations with $D=4$. Multiplying by $1/3!$ for three identical 
particles in the final state, we obtain 
\begin{equation}
\Gamma^{(0)}={1\over m^2}{(N_c^2-1)(N_c^2-4)\over 9N_c^2}(\pi^2-9)
\alpha_s^3,
\label{3s1-lh-0}
\end{equation}
\begin{equation}
\Gamma^{(2)}=-{{\bf p}^2\over m^4}{(N_c^2-1)(N_c^2-4)\over 108N_c^2}
(19\pi^2-132)\alpha_s^3.
\label{3s1-lh-2}
\end{equation}

In the case of the term proportional to ${\bf p}^4$, we must first
separate the IR singular parts in the matrix element
squared.\footnote{In Ref.~\cite{Petrelli:1998ge}, an alternative method
for dealing with the singular part was employed. The region of
integration was partitioned into three regions that are related by
interchange of the three gluon momenta. Only the region containing the
singularity at $x=0$ was retained, and the contribution from this region
was multiplied by three to obtain the complete result. The method that
we present in this paper has the advantage that the limits of
integration are simpler, and hence, the integrals are evaluated more
easily. Also, certain terms that cancel between the singular and
non-singular contributions in the method of Ref.~\cite{Petrelli:1998ge}
never appear in the present method.} These are given by
\begin{eqnarray}
{\tilde{\cal M}}_{\rm IR} &=&{{\bf p}^4\over  m^8}
\frac{(N_c^2-1)(N_c^2-4)}{N_c^2} \frac{128
\pi^3 \alpha_s^3 }{(3-2\epsilon)^3}
\left[(3-2\epsilon)(1-\epsilon) -
2(2-\epsilon) y(1-y)\right] \mu^{6\epsilon}\nonumber\\
&&\hbox{}\times
\left({1\over x^2}+{1\over (1-xy)^2}+{1\over [1-x(1-y)]^2}\right).
\label{IR-sing}
\end{eqnarray}
Integrating ${\tilde{\cal M}}_{\rm IR}$ over the phase space 
(\ref{xy-ps}) and multiplying by $1/3!$ for three identical particles in 
the final state, we obtain
\begin{subequations}
\label{3s1-lh-4-ir}
\begin{equation}
\Gamma_{\rm IR}^{(4)}=-{{\bf p}^4\over m^6}
\frac{(N_c^2-1)(N_c^2-4)}{N_c^2}\alpha_s^3                     
\frac{1}{\epsilon}\left(\frac{4\pi}{M^2}\right)^{2\epsilon}
\mu^{6\epsilon}
\frac{\Gamma^2(1-\epsilon)}{\Gamma^2(2-2\epsilon)}
{(1-\epsilon)^2(7-4\epsilon)\over (3-2\epsilon)^4}.
\end{equation}
Neglecting terms of order $\epsilon$, we may write this expression as
\begin{equation}
\Gamma_{\rm IR}^{(4)}=-{{\bf p}^4\over m^6}
\frac{(N_c^2-1)(N_c^2-4)}{N_c^2}\alpha_s^3\left[                     
\frac{7}{81\epsilon}\left(\frac{4\pi}{M^2}\right)^{2\epsilon}
\mu^{6\epsilon}
\frac{(1-\epsilon\gamma_E)\Gamma(1-\epsilon)}{\Gamma(2-2\epsilon)}
+\frac{44}{243}\right],
\end{equation}
\end{subequations}
where $\gamma_E$ is Euler's constant. After subtracting the IR singular
terms (\ref{IR-sing}) from the integrand, we can carry out the
phase-space integration over the remainder with $D=4$. Multiplying by
$1/3!$ for three identical particles in the final state, we obtain
\begin{equation}
\Gamma_{\rm finite}^{(4)}={{\bf p}^4\over m^6}
\frac{(N_c^2-1)(N_c^2-4)}{N_c^2}\alpha_s^3
\left[-\frac{3563}{2430}+\frac{1609}{6480}\pi^2 
\right].
\label{3s1-lh-4-finite}
\end{equation}
The complete decay width in full QCD of a ${}^3S_1$ $Q\bar Q$ state into light 
hadrons through order $v^4$ is then
\begin{equation}
\Gamma({}^3S_1\rightarrow {\rm LH})=\Gamma^{(0)}+\Gamma^{(2)}+
\Gamma^{(4)}_{\rm IR}+\Gamma^{(4)}_{\rm finite},
\label{3s1-lh-full}
\end{equation}
where the quantities on the right side are given in 
Eqs.~(\ref{3s1-lh-0}), (\ref{3s1-lh-2}), (\ref{3s1-lh-4-ir}), and 
(\ref{3s1-lh-4-finite}).

To determine the short-distance coefficients, we match these results
with the NRQCD expression for the decay width (\ref{Gpsilh}), evaluated
in the ${}^3S_1$ $Q\bar Q$ state. Since we have computed the full QCD
decay rate in order $\alpha_s^3$, we must evaluate each contributing
term in Eq.~(\ref{Gpsilh}) with accuracy $\alpha_s^3$. The coefficients
$F_8({}^1S_0)$, $F_8({}^3S_1)$, and $F_8({}^3P_J)$ are of order
$\alpha_s^2$. Therefore, we must evaluate the corresponding matrix
elements through order $\alpha_s$. We evaluate the matrix elements
corresponding to the unknown coefficients $F_1({}^3S_1)$, $G_1({}^3S_1)$,
$H_1^1({}^3S_1)$, and $H_1^2({}^3S_1)$ at order $\alpha_s^0$.

The color-octet matrix elements $\langle {}^3S_1| {\cal O}_8({}^1S_0)
|{}^3S_1\rangle$ and $\langle {}^3S_1| {\cal O}_8({}^3S_1) 
|{}^3S_1\rangle$ have a vanishing contribution at order $\alpha_s^0$ in 
the color-singlet $Q\bar Q$ state. The order-$\alpha_s^1$ contribution
comes from four diagrams in which a gluon connects an initial-state $Q$
or $\bar Q$ with a final-state $Q$ or $\bar Q$. The interaction of the
gluon with the $Q$ or $\bar Q$ cannot be of the ${\bf p}\cdot {\bf A}$
form, since that interaction changes the orbital angular momentum by one
unit. Any other NRQCD interaction must involve at least one power of the
gluon momentum. Hence, it is easy to see, by simple power counting
arguments, that the integration over the gluon momentum is ultraviolet 
(UV) power divergent. It therefore vanishes in dimensional regularization.

The color-octet matrix elements $\langle {}^3S_1| {\cal O}_8({}^3P_J)
|{}^3S_1\rangle$ also have a vanishing contribution at order
$\alpha_s^0$ in the color-singlet $Q\bar Q$ state. Again, the
order-$\alpha_s^1$ contribution comes from four diagrams in which a
gluon connects an initial-state $Q$ or $\bar Q$ with a final-state $Q$
or $\bar Q$. The contribution at leading order in $v$ arises from ${\bf 
p}\cdot {\bf A}$ interactions between the gluon and the $Q$ or $\bar Q$.
A straightforward computation yields
\begin{equation}
\langle {}^3S_1| {\cal O}_8({}^3P_J)|{}^3S_1\rangle
={{\bf p}^4\over m^2}{8(2J+1)C_F\over 81\pi}
\alpha_s\int_0^\infty {dk\over k}.
\end{equation}
This integral has logarithmic IR and UV divergences. Since it is scale 
invariant, it vanishes in dimensional regularization. It can be written 
as 
\begin{eqnarray}
\langle {}^3S_1| {\cal O}_8({}^3P_J)|{}^3S_1\rangle
&=&{{\bf p}^4\over m^2}{4(2J+1)C_F\over 81\pi}
{\mu^{2\epsilon}\over \mu_\Lambda^{2\epsilon}}\alpha_s\nonumber\\
&&\hbox{}\times\bigg\{\bigg[{1\over \epsilon_{\rm UV}}
+\log(4\pi)-\gamma_E\bigg]-\bigg[{1\over \epsilon_{\rm IR}}
+\log(4\pi)-\gamma_E\bigg]\bigg\},
\label{me-dim-reg}
\end{eqnarray}
where $\epsilon_{\rm UV}$ and $\epsilon_{\rm IR}$ are $(4-D)/2$, and
$\mu_\Lambda$ is the NRQCD renormalization scale. We renormalize the
expression (\ref{me-dim-reg}) in the modified minimal subtraction
($\overline{\rm MS}$) scheme by subtracting the contribution
proportional to $1/ \epsilon_{\rm UV}+\log(4\pi)-\gamma_E$. The
renormalized matrix element is
\begin{equation}
\langle {}^3S_1| {\cal O}_8({}^3P_J)|{}^3S_1\rangle_{\overline{\rm MS}}
=-{{\bf p}^4\over m^2}{4(2J+1)C_F\over 81\pi}
{\mu^{2\epsilon}\over \mu_\Lambda^{2\epsilon}}\alpha_s
\bigg[{1\over \epsilon}+\log(4\pi)-\gamma_E\bigg],
\label{me-renorm}
\end{equation}
where we have made the identification $\epsilon_{\rm IR}=\epsilon$.

Making use of these results for the matrix elements, we find that the 
decay width in NRQCD in order $\alpha_s^3$ is
\begin{subequations}
\label{gamma-nrqcd-3s1-ggg}
\begin{eqnarray}
\Gamma_{\rm NRQCD}({}^3S_1 \rightarrow {\rm LH})&=&
2N_c[(1/m^2)F_1({}^3S_1)
+({\bf p}^2/m^4) G_1({}^3S_1)
+({\bf p}^4/m^6) H^1_{\gamma\gamma}({}^1S_0)\nonumber\\
&&\qquad\hbox{}+({\bf p}^4/m^6) H^2_{\gamma\gamma}({}^1S_0)]
+({\bf p}^4/m^6)\sum_{J=0,1,2}c_JF_8({}^3P_J),
\end{eqnarray}
where the factor $2N_c$ in front of the square brackets comes from the 
color and spin factors for normalized heavy-quark states and
\begin{equation}
c_J=-{2(N_c^2-1)\over 81\pi N_c}(2J+1)
{\mu^{2\epsilon}\over \mu_\Lambda^{2\epsilon}}\alpha_s
\bigg[{1\over \epsilon}+\log(4\pi)-\gamma_E\bigg].
\end{equation}
\end{subequations}

The short-distance coefficients $F_8({}^3P_J)$ have been computed 
$D=4-2\epsilon$ dimensions in order $\alpha_s^2$ by Petrelli {\it et 
al.} \cite{Petrelli:1998ge}:
\begin{subequations}
\label{f8-3pj-d}
\begin{eqnarray}
F_8({}^3P_0)&=&18\pi B_F\alpha_s^2
\bigg({4\pi\over M^2}\bigg)^\epsilon \mu^{4\epsilon}
{\Gamma(1-\epsilon)\over \Gamma(2-2\epsilon)}
{1-\epsilon\over 3-2\epsilon},\\
F_8({}^3P_1)&=&0,\\
F_8({}^3P_2)&=&4\pi B_F\alpha_s^2
\bigg({4\pi\over M^2}\bigg)^\epsilon \mu^{4\epsilon}
{\Gamma(1-\epsilon)\over \Gamma(2-2\epsilon)}
{6-13\epsilon+4\epsilon^2\over (3-2\epsilon)(5-2\epsilon)}.
\end{eqnarray}
\end{subequations}
It follows that
\begin{equation}
\sum_{J=0,1,2}c_JF_8({}^3P_J)=-\frac{(N_c^2-1)(N_c^2-4)}{N_c^2}
\alpha_s^3\left[                     
\frac{7}{81\epsilon}
\left(\frac{4\pi\mu^3}{M \mu_\Lambda}\right)^{2\epsilon}
\frac{(1-\epsilon\gamma_E)\Gamma(1-\epsilon)}{\Gamma(2-2\epsilon)}
-{1\over 15}\right],
\label{f8-3pj-sum}
\end{equation}
where we have neglected terms of order $\epsilon$.

Using Eq.~(\ref{f8-3pj-sum}), we can compare the width in full QCD
[Eq.~(\ref{3s1-lh-full})] with the width in NRQCD
[Eq.~(\ref{gamma-nrqcd-3s1-ggg})] to compute the short-distance 
coefficients. As expected, the IR poles in $\epsilon$ cancel, and we 
obtain 
\begin{subequations}
\begin{eqnarray}
F_1({}^3S_1)&=&{(N_c^2-1)(N_c^2 - 4)\over N_c^3}{(\pi^2 - 9)\over 18}
\alpha_s^3,\\
G_1({}^3S_1)&=&{(N_c^2-1)(N_c^2-4)\over N_c^3}\bigg({11\over 18}-
{19\over 216}\pi^2\bigg)\alpha_s^3,\\
H_1^1({}^3S_1)+H_1^2({}^3S_1)&=&{(N_c^2-1)(N_c^2-4)\over 
N_c^3}\bigg(-{833\over 972}+{1609\over 12960}\pi^2+{7\over 81}
\log{2m\over\mu_\Lambda}\bigg)\alpha_s^3.\nonumber\\
\end{eqnarray}
\end{subequations}
Our result for $F_1({}^3S_1)$ agrees with that given in
Ref.~\cite{Mackenzie:1981sf}, and our result for $G_1({}^3S_1)$ agrees
with that given in Ref.~\cite{Keung:jb}. Our result for
$H_1^1({}^3S_1)+H_1^2({}^3S_1)$ is new.

\section{Discussion}

In this paper, we have computed short-distance coefficients for the 
decays of a ${}^1S_0$ heavy-quarkonium state to two photons and to light 
hadrons and the decays of a ${}^3S_1$ heavy-quarkonium state to a lepton 
pair and to light hadrons. Specifically, we have computed the 
coefficients of the operators whose matrix elements are of order $v^4$ 
and whose quantum numbers are those of the quarkonium state.

In our computation, we are able to obtain only the combinations
$H^1({}^{2S+1}L_J)+H^2({}^{2S+1}L_J)$, rather than the individual
coefficients $H^1({}^{2S+1}L_J)$ and $H^2({}^{2S+1}L_J)$, because the
corresponding operators, ${\cal Q}^1({}^{2S+1}L_J)$ and ${\cal
Q}^2({}^{2S+1}L_J)$, have identical matrix elements for on-shell heavy
quarks in the center-of-momentum frame. In order to obtain the values of
the individual coefficients, it would be necessary to consider matrix
elements of the operators ${\cal Q}^1({}^{2S+1}L_J)$ and ${\cal
Q}^2({}^{2S+1}L_J)$ in which the heavy $Q\bar Q$ interact with
additional quanta before reaching the annihilation vertex. Alternatively,
one could consider matrix elements of the operators ${\cal
Q}^3({}^{2S+1}L_J)$, which, as we have shown in
Appendix~\ref{app:relation}, are related to the operators ${\cal
Q}^1({}^{2S+1}L_J)$ and ${\cal Q}^2({}^{2S+1}L_J)$ through the equations
of motion.

In Tables I--IV, we show the numerical values of the short-distance
coefficients that appear through order $v^4$ for the decays that we
consider in this paper. For each coefficient, we take into account only
the contribution that is leading in $\alpha_s$. In each case, we
normalize the short-distance coefficients to the coefficient of the
operator whose matrix element is of leading order in $v$. In the third
column of each table, we use the velocity-scaling rules \cite{bbl} to
estimate the size of the matrix element of the operator that is
associated with each coefficient, relative to the size of the matrix
element of leading order in $v$. In the case of the color-octet
operators, we adopt the approach of Ref.~\cite{Petrelli:1998ge},
multiplying the velocity-scaling estimate by a factor $1/(2N_c)$ to
account for the relative spin and color normalizations of the
color-singlet and color-octet operators as we have defined them in this
paper.

\begin{table}[htb]
\begin{tabular}{lcc}
Coefficient&Value&Matrix Element\\
\hline
$F_{\gamma\gamma}({}^1S_0)$&$1$&$1$\\
$G_{\gamma\gamma}({}^1S_0)$&$-1.33$&$v^2$\\
$H_{\gamma\gamma}^1({}^1S_0)+H_{\gamma\gamma}^2({}^1S_0)$&$1.51$&$v^4$
\end{tabular}
\caption{Short-distance coefficients and estimates of sizes of corresponding 
matrix elements for the decay of a ${}^1S_0$ quarkonium state to two 
photons.}
\end{table}

\begin{table}[htb]
\begin{tabular}{lcc}
Coefficient&Value&Matrix Element\\
\hline
$F_1({}^1S_0)$&$1$&$1$\\
$G_1({}^1S_0)$&$-1.33$&$v^2$\\
$F_8({}^3S_1)$&$0.75n_f$&$v^3/(2N_c)$\\
$F_8({}^1S_0)$&$1.88$&$v^4/(2N_c)$\\
$F_8({}^1P_1)$&$1.13$&$v^4/(2N_c)$\\
$H_1^1({}^1S_0)+H_1^2({}^1S_0)$&$1.51$&$v^4$
\end{tabular}
\caption{Short-distance coefficients and estimates of sizes of corresponding 
matrix elements for the decay of a ${}^1S_0$ quarkonium state to light 
hadrons.}
\end{table}

\begin{table}[htb]
\begin{tabular}{lcc}
Coefficient&Value&Matrix Element\\
\hline
$F_{ee}({}^3S_1)$&$1$&$1$\\
$G_{ee}({}^3S_1)$&$-1.33$&$v^2$\\
$H_{ee}^1({}^3S_1)+H_{ee}^2({}^3S_1)$&$1.61$&$v^4$
\end{tabular}
\caption{Short-distance coefficients and estimates of sizes of corresponding 
matrix elements for the decay of a ${}^3S_1$ quarkonium state to 
a lepton pair.}
\end{table}

\begin{table}[htb]
\begin{tabular}{lcc}
Coefficient&Value&Matrix Element\\
\hline
$F_1({}^3S_1)$&$1$&$1$\\
$G_1({}^3S_1)$&$-5.32$&$v^2$\\
$F_8({}^1S_0)$&$11.64\pi/\alpha_s$&$v^3/(2N_c)$\\
$F_8({}^3S_1)$&$4.66n_f\pi/\alpha_s$&$v^4/(2N_c)$\\
$F_8({}^3P_0)$&$34.93\pi/\alpha_s$&$v^4/(2N_c)$\\
$F_8({}^3P_1)$&$2.26-6.90n_f+5.17n_f\log(2m/\mu_\Lambda)$&$v^4/(2N_c)$\\
$F_8({}^3P_2)$&$9.31\pi/\alpha_s$&$v^4/(2N_c)$\\
$H_1^1({}^3S_1)+H_1^2({}^3S_1)$
&\hbox{}\quad$7.62+1.79\log(2m/\mu_\Lambda)$\quad\hbox{}
&$v^4$
\end{tabular}
\caption{Short-distance coefficients and estimates of sizes of corresponding 
matrix elements for the decay of a ${}^3S_1$ quarkonium state to light 
hadrons.}
\end{table}

In the case of charmonium, $v^2\approx 0.3$ and $\alpha_s(m_c)\approx 
0.35$. Then, we see from Tables I--III that the convergence of the $v$ 
expansion is reasonable for the ${}^1S_0$ decays into two photons and 
into light hadrons and for the ${}^3S_1$ decay  into light hadrons. 

On the other hand, the coefficients in Table IV cast some doubt on the
convergence of the $v$ expansion in the case of the ${}^3S_1$ decay into
light hadrons. In the case of charmonium, all of the contributions of
higher order in $v$ are larger in magnitude than the order-$v^0$
contribution, with the exception of the $H_1^1({}^3S_1)+H_1^2({}^3S_1)$
contribution. The color-octet coefficients, other than $F_8({}^3P_1)$,
are enhanced by $\pi/\alpha_s$, relative to $F_1({}^3S_1)$, since the
corresponding color-octet Fock states can decay into two gluons or into
light-quark pairs, rather than into three gluons. In addition to this
enhancement, some of the coefficients of $\pi/\alpha_s$ are quite large.
However, one can, through a redefinition of the color-singlet operators,
incorporate the factors $1/(2N_c)$, which we have associated with the
matrix elements, into the short-distance coefficients
\cite{Petrelli:1998ge}. Then, aside from the $\pi/\alpha_s$ enhancement,
only $F_8({}^3P_0)$ is especially large. In the case of the
color-singlet coefficients, $G_1({}^3S_1)$ is quite large in magnitude
relative to $F_1({}^3S_1)$. However, the quantity
$H_1^1({}^3S_1)+H_1^2({}^3S_1)$ is not significantly larger in magnitude
than $G_1({}^3S_1)$, giving some hope that the $v$ expansion may
ultimately be well behaved.

The estimates of the sizes of the relativistic corrections strongly
suggest that, in order to carry out a meaningful phenomenological
analysis of $S$-wave quarkonium decays into light hadrons, one would
need to take into account contributions beyond leading order in $v$.
(For a further discussion of this point, see
Ref.~\cite{Maltoni:2000km}.) All of the contributions listed in Table~IV,
except for that of $F_8({}^3P_1)$, would be needed to achieve a
precision of better than 50\%.

Unfortunately, most of the required matrix elements are unknown.
However, the number of unknown quantities can be reduced drastically by
making use of the heavy-quark spin symmetry and the vacuum-saturation
approximation \cite{bbl}, although the accuracy of these approximations
is not always sufficient to allow a calculation of the decay rates
through relative order $v^4$.  Owing to the heavy-quark spin symmetry,
the matrix elements of ${\cal O}_1({}^1S_0)$, ${\cal P}_1({}^1S_0)$,
${\cal O}_8({}^3S_1)$, ${\cal O}_8({}^1S_0)$, ${\cal Q}_1^1({}^1S_0)$,
and ${\cal Q}_1^2({}^1S_0)$ in a ${}^1S_0$ state are equal to the matrix
elements of ${\cal O}_1({}^3S_1)$, ${\cal P}_1({}^3S_1)$, ${\cal
O}_8({}^1S_0)$, ${\cal O}_8({}^3S_1)$, ${\cal Q}_1^1({}^3S_1)$, and
${\cal Q}_1^2({}^3S_1)$ in a ${}^3S_1$ state, respectively, up to
corrections of relative order $v^2$. Also owing to the heavy-quark spin
symmetry, the matrix elements of the operators ${\cal O}_8({}^3P_J)$ in a
${}^3S_1$ state are equal $(2J+1)/9$ times the matrix element of ${\cal
O}_8({}^1P_1)$ in a ${}^1S_0$ state, up to corrections of relative order
$v^2$. According to the vacuum-saturation approximation, the matrix
elements of the operators for the electromagnetic decays are equal to
the matrix elements of the color-singlet hadronic-decay operators with
the same quantum numbers, up to corrections of relative order $v^4$. It
also follows from the vacuum-saturation approximation that the matrix
element of ${\cal Q}_1^1({}^{2S+1}S_J)$ is equal to the square of the
matrix element of ${\cal P}_1({}^{2S+1}S_J)$ divided by the matrix
element of ${\cal O}_1({}^{2S+1}S_J)$, up to corrections of order $v^4$.
However, the matrix element of ${\cal Q}_1^2({}^{2S+1}S_J)$ is not known
to be related to the others.

The matrix elements of ${\cal O}_1({}^{3}S_1)$ in the $J/\psi$ and
$\Upsilon$ states are known from phenomenology. The matrix elements of
${\cal O}_1({}^{3}S_1)$ and ${\cal P}_1({}^{3}S_1)$ in the $J/\psi$ and
$\Upsilon$ states have also been computed on the lattice \cite{bks},
although the lattice determinations of the matrix elements of ${\cal
P}_1({}^{3}S_1)$ are rather imprecise, owing to large uncertainties in
the perturbation series that relates the lattice and continuum matrix
elements. According to the Gremm-Kapustin relation \cite{Gremm:1997dq},
for dimensionally regulated matrix elements, the matrix element of
${\cal P}_1({}^{3}S_1)$ is equal to the matrix element of ${\cal
O}_1({}^{3}S_1)$ times $(M-2m_{\rm pole})/m$, up to corrections of
relative order $v^2$. Here, $M$ is the quarkonium mass, and $m_{\rm
pole}$ is the heavy-quark pole mass. The remaining unknown operator
matrix elements could, in principle, be determined in lattice numerical
simulations.

\begin{acknowledgments}

We wish to thank Eric Braaten and Jungil Lee for critical readings of
the manuscript. We also thank Jungil Lee for confirming our result for
$H^1_{ee}({}^3S_1)+H^2_{ee}({}^3S_1)$. We wish to thank G.~Peter Lepage
for a number of illuminating discussions. This work was supported by the
U.~S.~Department of Energy, Division of High Energy Physics, under
Contract No.~W-31-109-ENG-38.

\end{acknowledgments}

\appendix

\section{Relation between the operators of order $v^4$}
\label{app:relation}

In this appendix, we demonstrate that the operators ${\cal
Q}_1^i({}^1S_0)$ [Eq.~(\ref{1s0-ops})] are related to each other by the
equations of motion, as are the operators ${\cal Q}_1^i({}^3S_1)$
[Eq.~(\ref{3s1-ops})], the vacuum-saturated versions of the ${\cal
Q}_1^i({}^1S_0)$, and the vacuum-saturated versions of the ${\cal 
Q}_1^i({}^3S_1)$. We assume that these operators are integrated 
over all space-time, so that we can employ integration by parts in 
re-writing them.

We begin by considering the operator
\begin{equation}
{\cal Q}_1^1({}^1S_0)=\psi^\dagger(-\ihalf\tensor{\bf D})^2\chi
\chi^\dagger(-\ihalf\tensor{\bf D})^2\psi.
\end{equation}
Now,
\begin{eqnarray}
\chi^\dagger(-i\tensor{\bf D})^2\psi&=&
\chi^\dagger(-i\rp-g{\bf A}+i\lp-g{\bf A})(-i\rp-g{\bf A}+i\lp-g{\bf 
A})\psi\nonumber\\
&=&\chi^\dagger[2(-i\rp-g{\bf A})^2+2(i\lp-g{\bf A})^2-(i\rp+i\lp)^2]\psi
\nonumber\\
&=&\chi^\dagger[4m(i\roarrow\partial_0-gA_0)+4m(i\loarrow\partial_0-gA_0)
-(i\rp+i\lp)^2]\psi\nonumber\\
&=&[4im\roarrow\partial_0-(-i\rp)^2]\chi^\dagger\psi,
\end{eqnarray}
where we have used the equations of motion at leading order in $v$ in 
the third line. Furthermore, under integration by parts, which is
equivalent to energy-momentum conservation in momentum space,
\begin{equation}
[\psi^\dagger(-i\tensor{\bf D})^2\chi]
[4im\roarrow\partial_0-(-i\rp)^2]\chi^\dagger\psi
\rightarrow \{[-4mi\roarrow\partial_0-(i\rp)^2][\psi^\dagger
(-i\tensor{\bf D})^2\chi]\}\chi^\dagger\psi.
\label{int-parts}
\end{equation}

Let us focus on the first $Q\bar Q$ bilinear on the right of
Eq.~(\ref{int-parts}). It is
\begin{eqnarray}
[-4mi\roarrow\partial_0-(i\rp)^2][\psi^\dagger(-i\tensor{\bf D})^2\chi]
&=&\psi^\dagger\{-4m(i\loarrow\partial_0+gA_0)(-i\tensor{\bf D})^2
-4m(-i\tensor{\bf D})^2(i\roarrow\partial_0-gA_0)\nonumber\\
&&\qquad\hbox{}+4m[gA_0,(-i\tensor{\bf D})^2]
-4m[i\roarrow\partial_0,(-i\tensor{\bf D})^2]\nonumber\\
&&\qquad\hbox{}-(i\lp+i\rp)^2(-i\tensor{\bf D})^2\}\chi.
\end{eqnarray}
Then, using the equations of motion, we have
\begin{eqnarray}
[-4mi\roarrow\partial_0-(i\rp)^2][\psi^\dagger(-i\tensor{\bf D})^2\chi]
&=&\psi^\dagger\{2(i\lp-g{\bf A})^2(-i\tensor{\bf D})^2
+2(-i\tensor{\bf D})^2(-i\rp-g{\bf A})^2\nonumber\\
&&\qquad\hbox{}+4m[gA_0,(-i\tensor{\bf D})^2]
-4m[i\roarrow\partial_0,(-i\tensor{\bf D})^2]\nonumber\\
&&\qquad\hbox{}-(i\lp+i\rp)^2(-i\tensor{\bf D})^2\}\chi\nonumber\\
&=&\psi^\dagger\{(-i\tensor{\bf D})^4
+4m[(-iD_0),(-i\tensor{\bf D})^2]\}\chi\nonumber\\
&=&\psi^\dagger[(-i\tensor{\bf D})^4-8m(\tensor{\bf D}\cdot g{\bf E}
+g{\bf E}\cdot\tensor{\bf D})]\chi,
\end{eqnarray}
where, in arriving at the second equality, we have dropped some terms
proportional to $[\tensor{D}_i,\roarrow{D}_j]=-2ig\epsilon_{ijk}B_k$ that
are order $v^2$ relative to the terms that we have retained.

Thus, taking into account both $Q\bar Q$ bilinears, we have
\begin{equation}
{\cal Q}_1^1({}^1S_0)\rightarrow \psi^\dagger[(-\ihalf\tensor{\bf 
D})^4-(m/2)(\tensor{\bf D}\cdot g{\bf E}
+g{\bf E}\cdot\tensor{\bf D})]\chi\chi^\dagger\psi.
\end{equation}
Carrying out this procedure symmetrically on the left and right $Q\bar
Q$ bilinears of ${\cal Q}_1^1({}^1S_0)$, we conclude that, under the
equations of motion and integration by parts,
\begin{equation}
{\cal Q}_1^1({}^1S_0)\rightarrow {\cal Q}_1^2({}^1S_0)
+(m/2){\cal Q}_1^3({}^1S_0).
\end{equation}
A similar analysis in the spin-triplet case yields
\begin{equation}
{\cal Q}_1^1({}^3S_1)\rightarrow {\cal Q}_1^2({}^3S_1)
+(m/2){\cal Q}_1^3({}^3S_1).
\end{equation}
The vacuum-saturated versions of these relations, which are relevant to 
the electromagnetic decays are
\begin{eqnarray}
&&\langle{}^1S_0| \psi^\dagger(-\ihalf\tensor{\bf D})^2 
\chi |0\rangle
\langle 0| \chi^\dagger (-\ihalf \tensor{\bf D})^2 \psi 
|{}^1S_0\rangle\nonumber\\
&&\hbox{}\qquad\rightarrow
{\rm \, Re \,}\left[ \langle{}^1S_0|\psi^\dagger
\chi|0\rangle
\langle 0| \chi^\dagger (-\ihalf \tensor{\bf D})^4 \psi
|{}^1S_0\rangle \right]\nonumber\\
&&\hbox{}\qquad\phantom{\rightarrow \,}
+(m/2){\rm \, Re \,}\left[ \langle{}^1S_0| \psi^\dagger
\chi
|0\rangle \langle 0| \chi^\dagger 
(\tensor{\bf D}\cdot g{\bf E}+g{\bf E}\cdot\tensor{\bf D}) 
\psi |{}^1S_0\rangle \right]
\end{eqnarray}
and
\begin{eqnarray}
&&\langle{}^3S_1| \psi^\dagger \bm{\sigma}
(-\ihalf\tensor{\bf D})^2 \chi 
|0\rangle \cdot
        \langle 0| \chi^\dagger \bm{\sigma} (-\ihalf\tensor{\bf D})^2
\psi |{}^3S_1\rangle \nonumber\\
&&\hbox{}\qquad\rightarrow
{\rm \, Re \,}\left[\langle{}^3S_1|\psi^\dagger\bm{\sigma} 
\chi
|0\rangle \cdot
        \langle 0| \chi^\dagger \bm{\sigma}(-\ihalf \tensor{\bf D})^4  
\psi |{}^3S_1\rangle \right]\nonumber\\
&&\hbox{}\qquad\phantom{\rightarrow \,} 
+(m/2) {\rm \, Re \,}\left[ \langle{}^3S_1| \psi^\dagger
\bm{\sigma}\chi|0\rangle \cdot \langle 0| \chi^\dagger \bm{\sigma}
(\tensor{\bf D}\cdot g{\bf E}+g{\bf E}\cdot \tensor{\bf D})
\psi |{}^3S_1\rangle \right].
\end{eqnarray}

\section{Squared matrix elements for ${}^3S_1$ decay into light hadrons}
\label{app:sq-me}

In this appendix we give the expressions for the terms of order ${\bf
p}^{(0)}$, ${\bf p}^{(2)}$, and ${\bf p}^{(4)}$ in the square of the
matrix element for a ${}^3S_1$ $Q\bar Q$ state to decay into light
hadrons (three gluons). These terms are denoted by ${\tilde{\cal
M}}^{(0)}$, ${\tilde{\cal M}}^{(2)}$, and ${\tilde{\cal M}}^{(4)}$,
respectively. The quantity $m$ is the heavy-quark mass. The invariants
$s$, $t$, and $u$ are defined in Eq.~(\ref{stu}).

\begin{eqnarray}
{\tilde{\cal M}}^{(0)}&=&
{(N_c^2-1)(N_c^2-4)\over N_c^2}{2048\pi^3\alpha_s^3\over 3}
(16m^4s^2 - 8m^2s^3 + s^4 + 16m^4st - 12m^2s^2t + 2s^3t \br+ 
   16m^4t^2 - 12m^2st^2 + 3s^2t^2 - 8m^2t^3 + 2st^3 + t^4)\br
\times{1\over (4m^2 - s)^2(4m^2 - t)^2(s + t)^2};
\end{eqnarray}

\begin{eqnarray}
{\tilde{\cal M}}^{(2)}&=&
-{\bf p}^2{(N_c^2-1)(N_c^2-4)\over N_c^2}{8192\pi^3\alpha_s^3\over 9}
(768m^{10}s^2 - 256m^8s^3 - 48m^6s^4 + 24m^4s^5 \br - 2m^2s^6 + 
   512m^{10}st - 128m^8s^2t - 256m^6s^3t + 120m^4s^4t - 
   18m^2s^5t + s^6t \br + 768m^{10}t^2 - 128m^8st^2 - 256m^6s^2t^2 + 
   176m^4s^3t^2 - 37m^2s^4t^2 + 3s^5t^2 - 256m^8t^3 \br - 
   256m^6st^3 + 176m^4s^2t^3 - 44m^2s^3t^3 + 5s^4t^3 - 
   48m^6t^4 + 120m^4st^4 - 37m^2s^2t^4 \br + 5s^3t^4 + 24m^4t^5 - 
   18m^2st^5 + 3s^2t^5 - 2m^2t^6 + st^6)\br
\times{1\over m^2(4m^2 - s)^3(4m^2 - t)^3(s + t)^3};
\end{eqnarray}

\begin{eqnarray}
{\tilde{\cal M}}^{(4)}&=&
{\bf p}^4{(N_c^2-1)(N_c^2-4)\over N_c^2}{1024\pi^3\alpha_s^3\over 135}
(1966080m^{16}s^2 + 1409024m^{14}s^3 - 1445888m^{12}s^4 \br + 
   223232m^{10}s^5 + 47104m^8s^6 - 14080m^6s^7 + 880m^4s^8 + 
   1310720m^{16}st \br + 2588672m^{14}s^2t - 4034560m^{12}s^3t + 
   1176576m^{10}s^4t + 97536m^8s^5t \br - 86848m^6s^6t + 12352m^4s^7t - 
   552m^2s^8t + 1966080m^{16}t^2 + 2588672m^{14}st^2 \br - 
   3047424m^{12}s^2t^2 + 984064m^{10}s^3t^2 + 371456m^8s^4t^2 - 
   275264m^6s^5t^2  \br + 56032m^4s^6t^2 - 4688m^2s^7t^2 + 155s^8t^2 + 
   1409024m^{14}t^3 - 4034560m^{12}st^3 \br + 984064m^{10}s^2t^3 + 
   525312m^8s^3t^3 - 408960m^6s^4t^3 + 108304m^4s^5t^3 \br - 
   12452m^2s^6t^3 + 620s^7t^3 - 1445888m^{12}t^4 + 1176576m^{10}st^4 + 
   371456m^8s^2t^4 \br - 408960m^6s^3t^4 + 128640m^4s^4t^4 - 
   18468m^2s^5t^4 + 1240s^6t^4 + 223232m^{10}t^5 \br + 97536m^8st^5 - 
   275264m^6s^2t^5 + 108304m^4s^3t^5 - 18468m^2s^4t^5 + 
   1550s^5t^5 \br + 47104m^8t^6 - 86848m^6st^6 + 56032m^4s^2t^6 - 
   12452m^2s^3t^6 + 1240s^4t^6 \br - 14080m^6t^7 + 12352m^4st^7 - 
   4688m^2s^2t^7 + 620s^3t^7 + 880m^4t^8 - 552m^2st^8 \br + 
   155s^2t^8)
{1\over m^4(4m^2 - s)^4(4m^2 - t)^4(s + t)^4}.
\end{eqnarray}


\begin{thebibliography}{}

\bibitem{bbl}
G.~T.~Bodwin, E.~Braaten, and G.~P.~Lepage,
Phys.\ Rev.\ D {\bf 51}, 1125 (1995); 
{\bf 55}, 5853(E) (1997) 
[hep-ph/9407339].

\bibitem{Barbieri:1979be}
R.~Barbieri, E.~d'Emilio, G.~Curci, and E.~Remiddi,
Nucl.\ Phys.\ {\bf B154}, 535 (1979).

\bibitem{Hagiwara:1980nv}
K.~Hagiwara, C.~B.~Kim, and T.~Yoshino,
Nucl.\ Phys.\ {\bf B177}, 461 (1981).

\bibitem{Mackenzie:1981sf}
P.~B.~Mackenzie and G.~P.~Lepage,
Phys.\ Rev.\ Lett.\  {\bf 47}, 1244 (1981).

\bibitem{Keung:jb}
W.~Y.~Keung and I.~J.~Muzinich,
Phys.\ Rev.\ D {\bf 27}, 1518 (1983).

\bibitem{harris-brown} 
I.~Harris and L.~M.~Brown, Phys.\ Rev.\ {\bf 105}, 1656 (1957).

\bibitem{Barbieri:1975ki}
R.~Barbieri, R.~Gatto, R.~K\"ogerler, and Z.~Kunszt,
Phys.\ Lett.\ {\bf 57B}, 455 (1975).

\bibitem{Celmaster:1978yz}
W.~Celmaster,
Phys.\ Rev.\ D {\bf 19}, 1517 (1979).

\bibitem{Beneke:1997jm}
M.~Beneke, A.~Signer, and V.~A.~Smirnov,
Phys.\ Rev.\ Lett.\  {\bf 80}, 2535 (1998)
[hep-ph/9712302].

\bibitem{Petrelli:1998ge}
A.~Petrelli, M.~Cacciari, M.~Greco, F.~Maltoni, and M.~L.~Mangano,
Nucl.\ Phys.\ {\bf B514}, 245 (1998)
[hep-ph/9707223].

\bibitem{yang}
C.~N.~Yang, Phys.\ Rev.\ {\bf 77}, 242 (1950).

\bibitem{Kuhn:1979bb}
J.~H.~K\"uhn, J.~Kaplan, and E.~G.~Safiani,
Nucl.\ Phys.\ {\bf B157}, 125 (1979).

\bibitem{Guberina:1980dc}
B.~Guberina, J.~H.~K\"uhn, R.~D.~Peccei, and R.~R\"uckl,
Nucl.\ Phys.\ {\bf B174}, 317 (1980).

\bibitem{Mertig:an}
R.~Mertig, M.~Bohm, and A.~Denner,
Comput.\ Phys.\ Commun.\  {\bf 64}, 345 (1991).

\bibitem{Maltoni:2000km}
F.~Maltoni,
hep-ph/0007003.

\bibitem{bks}
G.~T.~Bodwin, S.~Kim, and D.~K.~Sinclair,
Nucl.\ Phys.\ B (Proc.\ Suppl.)  {\bf 34}, 434 (1994);
{\bf 42}, 306 (1995)
[hep-lat/9412011];
G.~T.~Bodwin, D.~K.~Sinclair, and S.~Kim,
Phys.\ Rev.\ Lett.\  {\bf 77}, 2376 (1996)
[hep-lat/9605023];
Int.\ J.\ Mod.\ Phys.\ A {\bf 12}, 4019 (1997)
[hep-ph/9609371];
Phys.\ Rev.\ D {\bf 65}, 054504 (2002)
[hep-lat/0107011].

\bibitem{Gremm:1997dq}
M.~Gremm and A.~Kapustin,
Phys.\ Lett.\ B {\bf 407}, 323 (1997)
[hep-ph/9701353].

\end{thebibliography}

\end{document}